\documentclass[journal,trans]{IEEEtran}
\usepackage{float}
\usepackage{soul}
\usepackage{mathtools}
\usepackage{epsfig,wrapfig}
\usepackage{epstopdf} 
\usepackage{caption}
\usepackage{subcaption}
\usepackage{fancyhdr}
\usepackage{psfrag}
\usepackage{lipsum}
\usepackage{xcolor}
\usepackage{multirow}
\usepackage{booktabs}
\usepackage{rotfloat}
\usepackage{cite}
\usepackage{amsmath}
\usepackage{cite}

\definecolor{cadmiumgreen}{rgb}{0.0, 0.42, 0.24}
\DeclareUnicodeCharacter{2212}{-}

\newcommand{\D}{\partial}
\newcommand{\B}{\mathbf}
\newcommand{\tx}{\text}

\newcommand{\eps}{\epsilon}

\newcommand{\mt}{\mathrm}

\makeatletter
\newcommand{\notparallel}{%
  \mathrel{\mathpalette\not@parallel\relax}%
}
\newcommand{\pati}[1]{\reversemarginpar\hspace{0pt}\marginpar{\hspace{-2.2cm}\begin{minipage}{2.8cm}\centering\vspace{\baselineskip}\textcolor{magenta}{\textit{\phantom{#1}}}\end{minipage}}}
\ifCLASSINFOpdf
\else
\fi
\begin{document}

\title{Generalized FDTD Scheme \\for Moving Electromagnetic Structures \\with Arbitrary Space-Time Configurations}

\author{\IEEEauthorblockN{Amir Bahrami,
Zo{\'e}-Lise Deck-L{\'e}ger,
Zhiyu Li and Christophe Caloz,~\IEEEmembership{Fellow,~IEEE}}

}

\IEEEtitleabstractindextext{%
\begin{abstract}
We present a generalized FDTD scheme to simulate moving electromagnetic structures with arbitrary space-time configurations. This scheme is a local adaptation and 2+1-dimensional extension of the uniform and 1+1-dimensional scheme recently reported in~\cite{deck2022yeecell}. The local adaptation, which is allowed by the inherently matched nature of the generalized Yee cell to the conventional Yee cell, extends the range of applicability of the scheme in~\cite{deck2022yeecell} to moving structures that involve \emph{multiple and arbitrary velocity profiles} while being fully compatible with conventional absorbing boundary conditions and standard treatments of medium dispersion. We show that a direct application of the conventional FDTD scheme predicts qualitatively correct spectral transitions but quantitatively erroneous scattering amplitudes, we infer from this observation generalized, hybrid -- physical and auxiliary (non-physical) -- fields that automatically satisfy moving boundary conditions in the laboratory frame, and accordingly establish local update equations based on the related Maxwell's equations and constitutive relations. We subsequently provide a detailed stability analysis with a generalization of the Courant criterion to the dynamic regime. We finally validate and illustrate the proposed method by several representative examples. The proposed scheme fills an important gap in the open literature on computational electromagnetics and offers an unprecedented, direct solution for moving structures in commercial software platforms.
\end{abstract}

}
\maketitle
\IEEEdisplaynontitleabstractindextext
\IEEEpeerreviewmaketitle


\section{Introduction}\label{sec:Intro}

\pati{Moving Electromagnetic Structures} 
The introduction of motion into electromagnetic structures represents a fundamental extension of stationary electromagnetics. The motion may involve either moving matter, such as rotating dielectrics and accelerated charges, or moving perturbations, such as fluid and elastic waves. The structures involving the latter type of motion have been recently generalized to Space-Time Engineered-Modulation (GSTEM) metamaterials, or GSTEMs for short, which encompass a virtually unlimited diversity of space-time configurations (e.g., co- or contra-directional, single- or multiple-interface, harmonic, step or gradient, 1+1, 2+1 or 3+1 dimensional (D), uniform or accelerated, classical or quantum, etc.)~\cite{caloz2022gstem}, and hence dramatically extend the physics diversity and application potential of previous moving electromagnetic structures~\cite{bradley1729iv,Fresnel1818,doppler1903ueber,Fizeau1851,Rontgen_1888,Minkowski_1908,tien1958parametric,cassedy1963dispersion,yeh1965,cassedy1967dispersion,kong_1968_wave,tanaka1978relativistic,leonhardt1999optics,Reed_2003_color,Lurie_Springer_2007,Biancalana_2007_dynamics,philbin_2008_fiber,Yu_2009_opticalisolation,belgiorno_2010_hawking,faccio2012optical,faccio2013analogue,Deck_2018_wave,Shaltout_Science_2019,Deck_APH_10_2019,caloz2019spacetime1,caloz2019spacetime2,huidobro2019fresnel,Gaafar_2019_front,huidobro2021homogenization,Li_PRB_03_2023,bahrami2023astem}. \emph{Moving electromagnetic structures} represent thus a field of science and technology that is more vibrant than ever.

\pati{Modeling State-of-the-Art and Gap}
However, surprisingly, no general numerical tool was available for simulating electromagnetic moving structures until the very recent \emph{Finite-Difference Time-Domain (FDTD)}~\cite{yee1966numerical,taflove2005computational} scheme reported in~\cite{deck2022yeecell}, which is based on a generalized Yee cell with hybrid physical and nonphysical auxiliary fields that automatically satisfy moving boundary conditions. Previous related state-of-the-art was limited to two other FDTD approaches, one restricted to non-penetrable objects~\cite{Harfoush1990,Harfoush1989} and one implying cumbersome Lorentz frame transformations~\cite{iwamatsu2009,zheng2016,zhao_2018,Bahrami_MTM_09_2022}. The scheme in~\cite{deck2022yeecell} extends the capabilities of~\cite{Harfoush1990,Harfoush1989} and avoids the issues of~\cite{iwamatsu2009,zheng2016,zhao_2018,Bahrami_MTM_09_2022} by offering an efficient treatment of moving penetrable media, including gradient structures and metamaterials. However, it still suffers from major limitations, including the incapability to handle moving structures that involve different velocities or/and nonuniform (or accelerated) space-time configurations, due to the uniform nature of the generalized Yee cell, and general incompatibility with conventional absorbing boundary conditions~\cite{mur1981absorbing,berenger1993} and impossibility to model dispersive media, due to the hybrid nature of the update fields.

\pati{Contribution}
We present here an elaborated version of the FDTD scheme in \cite{deck2022yeecell} that overcomes all of these issues, i.e., that can handle \emph{arbitrary space-time configurations} while being compatible with conventional absorbing boundary conditions and being capable to account for medium dispersion. First, we demonstrate and explain the failure of the conventional FDTD scheme to model general moving electromagnetic structures (Sec.~\ref{sec:Failure}). Then, we state the well-known moving boundary conditions and show how these boundary conditions lead to the ``generalized'' Maxwell's equations that form the basis of the generalized Yee-cell FDTD scheme in~\cite{deck2022yeecell} (Sec.~\ref{sec:Aux}). Next, we show that the Yee-cell uniformity inherent to that scheme causes the aforementioned limitations and we propose an alternative, \emph{local} scheme that overcomes these limitations and hence offers a solution for modeling arbitrary space-time configurations (Sec.~\ref{sec:Local_Treat}). We subsequently provide the related 2+1D\footnote{The update equations in~\cite{deck2022yeecell} are restricted to 1+1 dimensions.} update equations (Sec.~\ref{sec:Disc_Eq}) and a detailed stability analysis for the proposed scheme (Sec.~\ref{sec:stability_anal}). Finally, we validate and illustrate the proposed method by four examples: a space-time interface under oblique incidence, a space-time wedge, a space-time accelerated interface, which combines the features of arbitrary space-time configurations, and a curved space-time interface (Sec.~\ref{sec:Validation}). Finally, we close the paper with a few concluding remarks (Sec.~\ref{sec:Conclusion}).

\section{Failure of the Conventional FDTD Scheme}\label{sec:Failure}

\pati{Choice of the Uniform Space-Time Discontinuity}
A moving structure, as shown in~\cite{caloz2022gstem}, may always be decomposed into a succession or mixture of moving discontinuities. A \emph{moving discontinuity} -- or \emph{space-time discontinuity} -- which may also be seen as a \emph{moving interface between two media}, is therefore the building block of any moving structure. We shall show here that the conventional FDTD scheme fails to model such a moving discontinuity.

\begin{figure}[ht!]
    \centering
    \includegraphics[width=8.6cm]{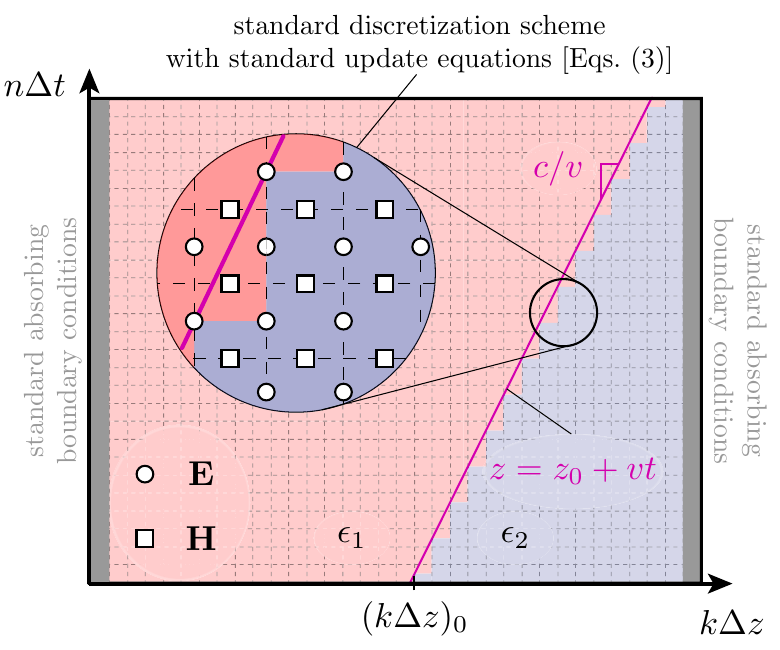}
    \caption{Naive application of the conventional FDTD scheme, with its standard update equations and absorbing boundary conditions, to model a space-time discontinuity (moving at the velocity $v$) between two media of different permittivities ($\epsilon_1$ and $\epsilon_2$): simple rotation of the pure-space, vertical boundary into an oblique boundary in the simulation setup.}\label{fig:conv_FDTD_approach}
\end{figure}

\pati{Conventional FDTD Naive Approach}
It may \emph{a priori} seem, under the perspective of the space-time (or Minkowski) diagram represented in Fig.~\ref{fig:conv_FDTD_approach}, that the conventional FDTD scheme, with its standard update  equations and absorbing boundary conditions, could straightforwardly apply to the problem of a space-time discontinuity. In that perspective, the space-time discontinuity is indeed just a rotated, oblique version of the routine pure-space (or stationary), vertical discontinuity between two media, and it can easily be specified as such in the parametric setup of the simulation. Note that both the space step ($\Delta z$) and the time step ($\Delta t$) may have to be decreased compared to the pure-space discontinuity case for maintaining the same level of accuracy due to wavelength or/and period compression induced by Doppler or/and index contrast effects~\cite{caloz2019spacetime2}.  
%

\pati{Testing of the Naive Approach}
Let us test the strategy outlined in the previous paragraph. Discretizing Maxwell's equations,
\begin{subequations}\label{eq:Maxwell}
\begin{equation}\label{eq:Maxwell_E}
\nabla\times\B{E}=-\frac{\D \B{B}}{\D t}
\end{equation}
and
\begin{equation}\label{eq:Maxwell_H}
\nabla\times\B{H}=\frac{\D \B{D}}{\D t}
\end{equation}
\end{subequations}
with the constitutive relations
\begin{subequations}\label{eq:Const_Iso}
\begin{equation}\label{eq:Const_Iso_eps}
\B{D}=\epsilon\B{E}
\end{equation}
and
\begin{equation}\label{eq:Const_Iso_mu}
\B{B}=\mu\B{H}
\end{equation}
\end{subequations}
in the usual way~\cite{taflove2005computational} yields, in two dimensions ($y$ and $z$, assuming $\partial/\partial x=0$) and for s-polarization\footnote{The sequel of the paper is restricted to s-polarization. The case of p-polarization can be treated in an analogous manner.},
\begin{subequations}\label{eq:Conv_FDTD}
\begin{equation}\label{eq:Conv_FDTD_By}
B_y|^{n}_{k+\frac{1}{2},i+\frac{1}{2}}=B_y|^{n-1}_{k+\frac{1}{2},i+\frac{1}{2}}
-\frac{\Delta t}{\Delta z}\left(E_x|^{n-\frac{1}{2}}_{k+1,i}- E_x|^{n-\frac{1}{2}}_{k,i}\right),
\end{equation}

\begin{equation}\label{eq:Conv_FDTD_Hy}
H_y|_{k+\frac{1}{2},i+\frac{1}{2}}^{n}=\frac{B_y|_{k+\frac{1}{2},i+\frac{1}{2}}^{n}}{\mu|_{k+\frac{1}{2},i+\frac{1}{2}}^{n}},
\end{equation}

\begin{equation}\label{eq:Conv_FDTD_Bz}
B_z|^{n}_{k+\frac{1}{2},i+\frac{1}{2}}=B_z|^{n-1}_{k+\frac{1}{2},i+\frac{1}{2}}
+\frac{\Delta t}{\Delta y}\left(E_x|^{n-\frac{1}{2}}_{k,i+1}- E_x|^{n-\frac{1}{2}}_{k,i}\right),
\end{equation}

\begin{equation}\label{eq:Conv_FDTD_Hz}
H_z|_{k+\frac{1}{2},i+\frac{1}{2}}^{n}=\frac{B_z|_{k+\frac{1}{2},i+\frac{1}{2}}^{n}}{\mu|_{k+\frac{1}{2},i+\frac{1}{2}}^{n}},
\end{equation}

\begin{equation}\label{eq:Conv_FDTD_D}
\begin{split}
 D_x|^{n+\frac{1}{2}}_{k,i}=D_x|^{n-\frac{1}{2}}_{k,i}
&+\frac{\Delta t}{\Delta y}\left(H_z|^{n}_{k+\frac{1}{2},i+\frac{1}{2}}-H_z|^{n}_{k+\frac{1}{2},i-\frac{1}{2}}\right)
\\&-\frac{\Delta t}{\Delta z}\left(H_y|^{n}_{k+\frac{1}{2},i+\frac{1}{2}}-H_y|^{n}_{k-\frac{1}{2},i+\frac{1}{2}}\right)
\end{split}
\end{equation}
and
\begin{equation}\label{eq:Conv_FDTD_E}
E_x|^{n+\frac{1}{2}}_{k,i}=\frac{D_x|^{n+\frac{1}{2}}_{k,i}}{\epsilon|^{n+\frac{1}{2}}_{k,i}},
\end{equation}
\end{subequations}
where the permittivity may be written, according to Fig.~\ref{fig:conv_FDTD_approach}, as
\begin{equation}\label{eq:Conv_FDTD_Const_eps}
\epsilon|^{n+\frac{1}{2}}_{k,i}=\epsilon(k\Delta z,n\Delta t)=
  \begin{cases}
    \epsilon_1~\mathrm{if}~k\Delta z \leq (k\Delta z)_0 +vn\Delta t,\\
    \epsilon_2~\mathrm{if}~k\Delta z > (k\Delta z)_0 +vn\Delta t,
  \end{cases}
\end{equation}
where $k$, $i$, $n$ and $(k\Delta z)_0$ are the spatial index along the $z$ direction, the spatial index along the $y$ direction, the temporal index and the initial position of the interface on the spatial grid, respectively.

%

\begin{figure}   
    \centering
    \includegraphics[width=8.6cm]{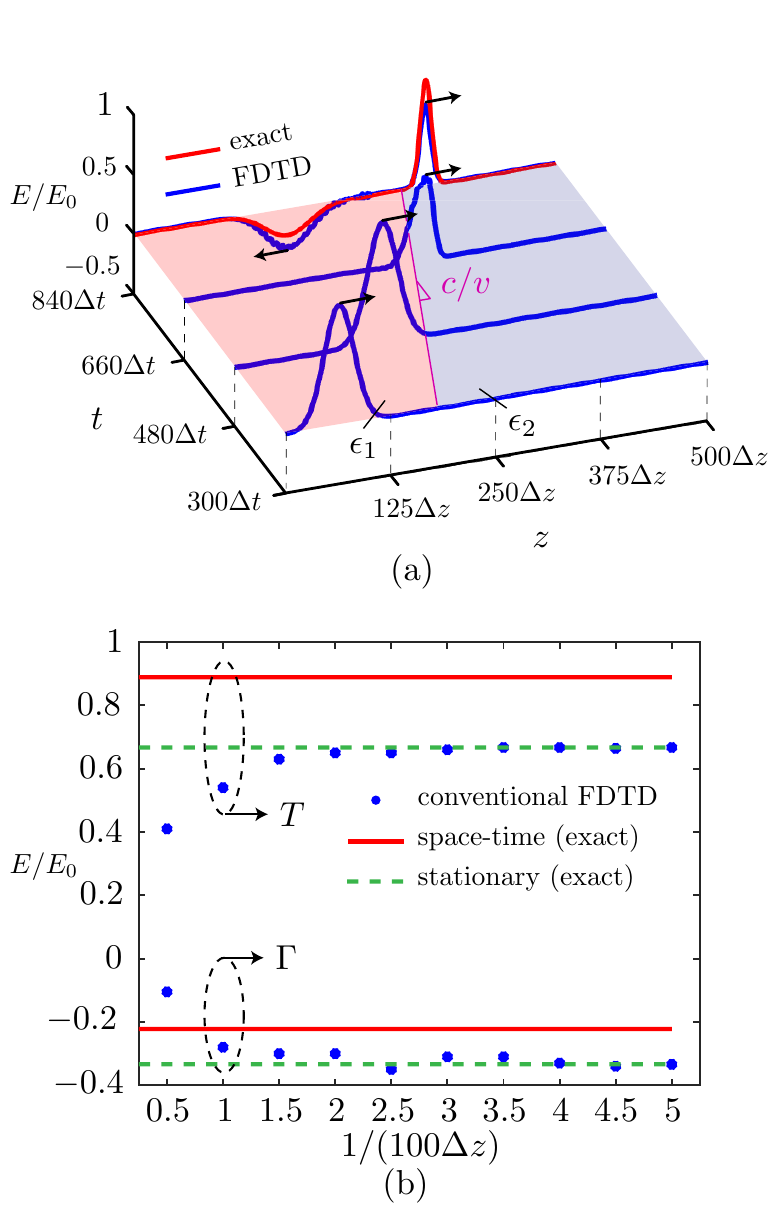}
    \caption{Failure of the conventional FDTD scheme [Fig.~\ref{fig:conv_FDTD_approach} and Eqs.~\eqref{eq:Conv_FDTD} with~\eqref{eq:Conv_FDTD_Const_eps}] to model a space-time discontinuity, here for the parameters $\epsilon_1=1$, $\epsilon_2=4$, $v=0.2c$ and $\Delta t=\Delta z/(2c)$, and for the electric field Gaussian pulse excitation $E=E_0e^{-(t-T_0)^2/\tau^2}$ with $T_0=200\Delta t$ and $\tau=40\Delta t$. (a)~Pulse evolution in space ($z$) and time ($t$). (b)~Reflection ($\Gamma$) and transmission ($T$) coefficients, measured from the peaks of the scattered pulses, versus increasing mesh density, $1/\Delta z$.}\label{fig:conv_FDTD_issue_illustr}
\end{figure}

\pati{Demonstration of Failure}
Figure~\ref{fig:conv_FDTD_issue_illustr} compares results obtained by the just described approach with exact results given by analytical formulas that are provided in Appendix~\ref{sec:App_USTEM} [Eqs.~\eqref{eq:Uni_Scat_Coeff}]. Figure~\ref{fig:conv_FDTD_issue_illustr}(a) shows that the FDTD-computed scattered pulse waveform strongly deviates from the exact result, which points to a basic malfunction of the scheme. To check whether this discrepancy might not just be an artifact of under-sampling, Fig.~\ref{fig:conv_FDTD_issue_illustr}(b) plots the scattering (reflection and transmission) coefficients for the pulse versus mesh density. It shows that the result in Fig.~\ref{fig:conv_FDTD_issue_illustr}(a) had reached proper convergence and quantitatively confirms the error ($24.98\%$ for transmission and $50.31\%$ for reflection) that was qualitatively observed in Fig.~\ref{fig:conv_FDTD_issue_illustr}(a), hence revealing a fundamental failure of the scheme to model a space-time discontinuity.

\pati{Explanation of the Issue}
Strangely, the coefficients in Fig.~\ref{fig:conv_FDTD_issue_illustr} converge to the \emph{stationary} instead of the space-time exact values! Why would this be the case? This question might be answered by considering that, as shown in the inset of Fig.~\ref{fig:conv_FDTD_approach}, the fields common to the two media, i.e., the fields that are exactly positioned at the interface between these media and are hence the fields that are forced to be continuous there, are the tangential $\mathbf{E}$ and $\mathbf{H}$ fields. Since the continuity of such fields is the proper boundary condition for a stationary interface (and not to a space-time interface!)~\cite{Jackson_1998}, the  erroneous result is eventually not surprising. The scheme properly delineates the space-time discontinuity in the computational space (Fig.~\ref{fig:conv_FDTD_approach}), which produces qualitatively correct spectral transformations (red-shift, wider-pulse reflection, due to contra-directional scattering and blue-shift, narrower-pulse, transmission, due to transmission from a rarer to a denser medium)~\cite{caloz2019spacetime2}, but, enforcing stationary (or instantaneous) instead of moving field continuity conditions, it fails to provide the correct scattering coefficients.
      
\pati{Inadequacy of Commercial Software} 
An important conclusion to be drawn from the observations done in this section is that currently available \emph{commercial FDTD software platforms} are not capable of simulating space-time discontinuities, and hence electromagnetic moving media. Such tools can, of course, handle pure-space discontinuities (which correspond to the stationary limit of space-time discontinuities~\cite{caloz2022gstem}), as they routinely do, since the related required continuity of the tangential $\mathbf{E}$ and $\mathbf{H}$ fields~\cite{caloz2019spacetime2} is embedded in the spatial evolution of discretized Maxwell's curl equations~\eqref{eq:Conv_FDTD_By}, \eqref{eq:Conv_FDTD_Bz} and~\eqref{eq:Conv_FDTD_D}, as shown in Appendix~\ref{sec:App_Natur_BCs}. In fact, they can \emph{also} handle \emph{pure-time discontinuities} (which correspond the instantaneous limit of space-time discontinuities~\cite{caloz2022gstem}) (e.g., \cite{Pena_2020_tempcoating,Pacheco_2020_temporalaming}), because the related required continuity of the tangential $\mathbf{D}$ and $\mathbf{B}$ fields~\cite{caloz2019spacetime2} is embedded in the \emph{temporal} evolution of the same equations, as also shown in Appendix~\ref{sec:App_Natur_BCs}. However, the discretized Maxwell's equations include no provision for satisfying the continuity conditions corresponding to space-time discontinuities beyond these two particular cases.

\section{Hybrid-Field Maxwell's Equations}\label{sec:Aux}

\pati{Proper Moving Boundary Conditions}
A correct scheme for modeling space-time discontinuities must naturally enforce \emph{the continuity of the corresponding fields} at the interface between the two media that forms the discontinuity, consistently with the well-known moving boundary conditions~\cite{pauli1981theory,bladel1984}
\begin{subequations}\label{eq:cont}
\begin{equation}\label{eq:Estarcont}
 \hat{\mathbf{n}}\times(\B{E}_2^*-\B{E}_1^*)=0,
\end{equation}
\begin{equation}\label{eq:Hstarcont}
  \hat{\mathbf{n}}\times(\B{H}_2^*-\B{H}_1^*)=\mathbf{J}_\tx{s},
\end{equation}
\begin{equation}\label{eq:Dcont}
     \hat{\mathbf{n}}\cdot(\B{D}_2-\B{D}_1)= \rho_\tx{s},
\end{equation}
\begin{equation}\label{eq:Bcont}
     \hat{\mathbf{n}}\cdot(\B{B}_2-\B{B}_1)= 0,
\end{equation}
\end{subequations}
with
\begin{equation}\label{eq:star}
    \B{E}^*=\B{E}+\mathbf{v}\times\B{B}
    \quad\text{and}\quad
    \B{H}^*=\B{H}-\mathbf{v}\times\B{D},
\end{equation}
where $1$ and $2$ label the media at the two sides the interface, $\B{J}_\tx{s}$ and $\rho_\tx{s}$ are the usual surface current and charge densities, respectively, $\hat{\mathbf{n}}$ is the unit vector normal the interface and pointing towards medium~1, and $\B{v}$ is the velocity of the interface, which is typically but not necessarily perpendicular to it~\cite{deck2022yeecell}.

\pati{\cite{deck2022yeecell}'s Scheme with Hybrid-Field Maxwell's Equations}
Equations~\eqref{eq:cont} and~\eqref{eq:star} reveal that the fields that are continuous at a (charge/current-less) space-time discontinuity are neither the $\B{E}$ and $\B{H}$ fields, nor the $\B{D}$ and $\B{B}$ fields, but the starred fields in~\eqref{eq:star}. This consideration inspired us to establish in~\cite{deck2022yeecell} a \emph{generalized Yee cell} with the usual, physical fields $\B{E}$ and $\B{H}$ being replaced by the \emph{hybrid fields} $\B{E}^*=\B{E}+\mathbf{v}\times\B{B}$ and $\B{H}^*=\B{H}-\mathbf{v}\times\B{D}$, which include the auxiliary, unphysical terms $\mathbf{v}\times\B{B}$ and $\mathbf{v}\times\B{D}$ in addition to the usual, physical fields, for an automatic satisfaction of the moving boundary conditions. The generalized Yee cell corresponds then to the \emph{hybrid-field Maxwell's equations} obtained by inserting Eqs.~\eqref{eq:star} into Eqs.~\eqref{eq:Maxwell}, viz.,
\begin{subequations}\label{eq:Maxwell_Aux}
\begin{equation}
\nabla\times\B{E}^*=-\frac{\partial\B{B}}{\partial t}+\nabla\times(\B{v}\times\B{B})
\end{equation}
and
\begin{equation}
\nabla\times\B{H}^*=\frac{\partial\B{D}}{\partial t}-\nabla\times(\B{v}\times\B{D}),
\end{equation}
\end{subequations}
and to the \emph{hybrid-field constitutive relations} 
obtained by inserting Eqs.~\eqref{eq:star} into Eqs.~\eqref{eq:Const_Iso},
\begin{subequations}\label{eq:Const_Aux}
\begin{equation}\label{eq:Const_Aux_E}
    \B{E}^*=\epsilon^{-1}\cdot \B{D}+\B{v}\times \B{B}
\end{equation}
and
\begin{equation}\label{eq:Const_Aux_H}
    \B{H}^*=\mu^{-1}\cdot \B{B}-\B{v}\times \B{D},
\end{equation}
\end{subequations}
which may be straightforwardly extended to bianisotropic relations\footnote{\label{fn:matter_motion}Such an extension is necessary for \emph{moving-matter} -- as opposed to \emph{moving-perturbation} -- structures, because matter, even when isotropic at rest, takes a particular form of bianisotropy when moving, due to related magneto-dielectric coupling~\cite{Rontgen_1888,pauli1981theory,kong2008theory,Deck_Photon_04_2021} . In this paper, we restrict ourselves, for the sake of simplicity, to moving-perturbation structures. However, the proposed scheme straightforwardly applies to the case of moving-matter structures, which is just unessentially complicated by the tensorial nature of the bianisotropic parameters.}.

\section{Local Treatment of Moving Boundaries}\label{sec:Local_Treat}

\pati{Principle of the~\cite{deck2022yeecell} FDTD Scheme}
The FDTD scheme presented in~\cite{deck2022yeecell} consists in first specifying the space-time constitutive parameters in the parametric setup of the simulation (as in the diagram of Fig.~\ref{fig:conv_FDTD_approach}), then running the FDTD leapfrog algorithm with the aforementioned generalized Yee cell, which involves the generalized, hybrid (globally unphysical) fields $\B{E}^*$ and $\B{H}^*$, and finally computing the physical fields $\B{E}$ and $\B{H}$ by inverting Eq.~\eqref{eq:Const_Aux} as \mbox{$\B{E}=\epsilon^{-1}\cdot\B{D}=\B{E}^*-\B{v}\times\B{B}$} and \mbox{$\B{H}=\mu^{-1}\cdot\B{B}=\B{H}^*+\B{v}\times\B{D}$}. It allows to simulate \emph{single-velocity} space-time interface, slab, crystal and gradient structures.

\pati{Limitations of the \cite{deck2022yeecell} due to Uniformity}
However, that scheme uses a generalized Yee cell that is, as the conventional Yee cell, \emph{uniform}, i.e., that is the same across the entire computational domain, and that uniformity restricts it to moving structures that involve a unique velocity, namely the velocity $\B{v}$ in Eqs.~\eqref{eq:Maxwell_Aux} and~\eqref{eq:Const_Aux}. This represents a major limitation, which prevents, for instance, the simulation of moving structures involving multiple discontinuities of different velocities, such as space-time wedges~\cite{balazs_1961_solution,Felsen_1970_wave,fante1971transmission} (example in Sec.~\ref{sec:ST_wedge} with velocities $\B{v}_\mathrm{I}$ and $\B{v}_\mathrm{II}$) and time-varying discontinuities, such as accelerated interfaces (example in Sec.~\ref{sec:acc_interf} with time-varying velocity, $\B{v}(t)$)~\cite{bahrami2023astem}. Moreover, the non-physicality of the hybrid fields across the entire computational domain implies a general incompatibility with standard absorbing conditions, such as Perfectly Matched Layers~\cite{berenger1993}, and the impossibility to properly account for (physical) medium dispersion using standard related techniques~\cite{taflove2005computational}.

\begin{figure}[h!]
    \centering
    \includegraphics[width=8.6cm]{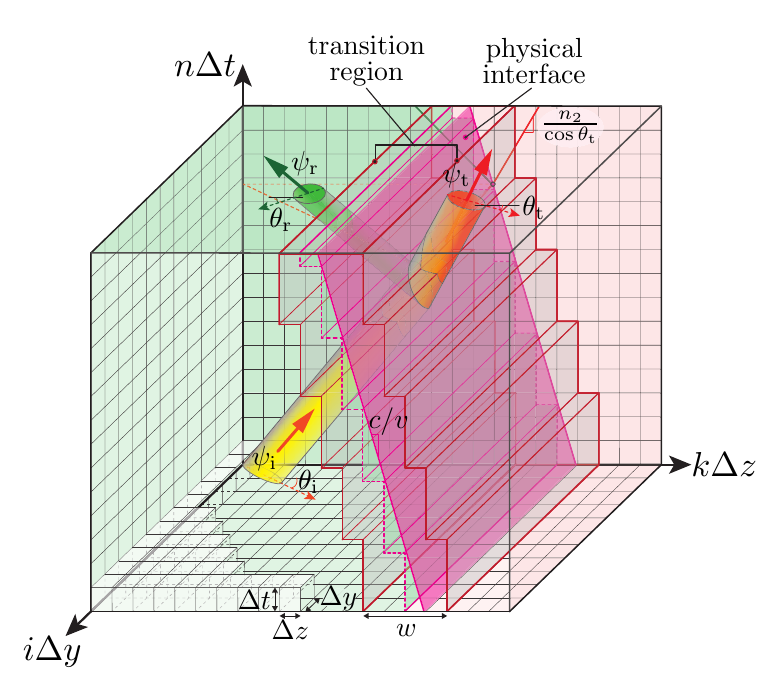}
    \caption{Local treatment of the generalized Yee-cell scheme, overcoming the limitations of the uniform generalized Yee-cell scheme in~\cite{deck2022yeecell}, for oblique light incidence (2+1D problem) on a contra-directionally moving interface, with obliquity being represented by the oblique space-time trajectory of a light pulse (with its incident, reflected and transmitted parts).}\label{fig:3D_Grid}
\end{figure}

\begin{figure}[h!]   
    \centering
    \includegraphics[width=8.6cm]{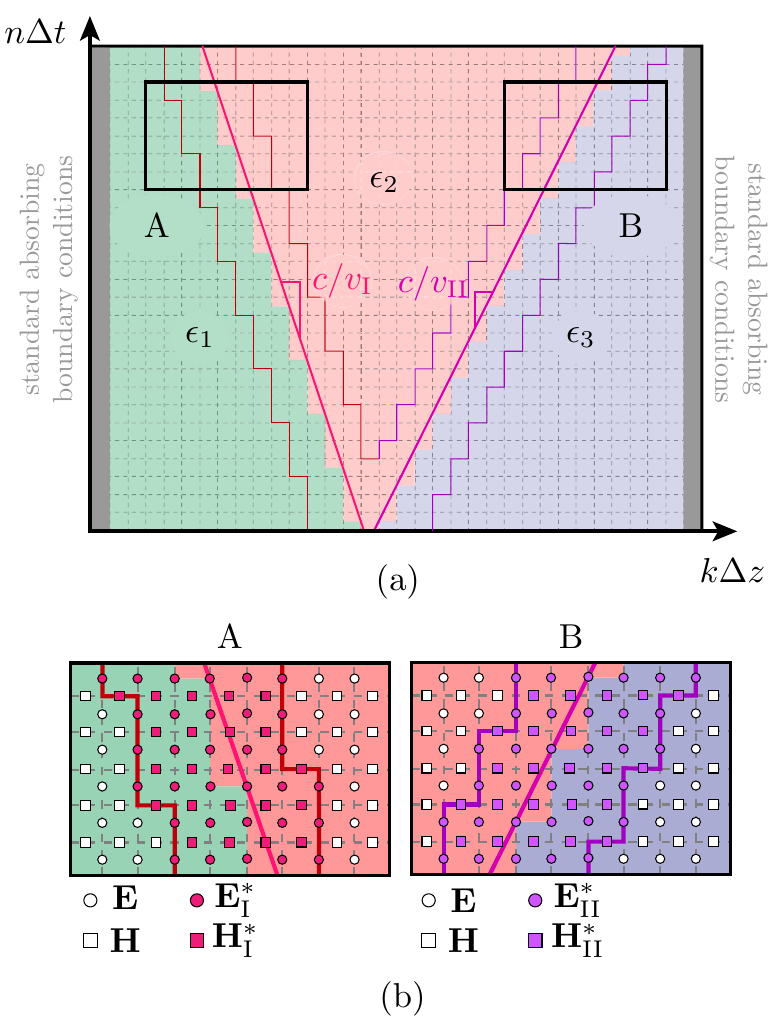}
    \caption{$zt$-plane (1+1D) projection of Fig.~\ref{fig:3D_Grid} with an additional interface for greater space-time structural generality. (a)~Global view of the structure, with its two interfaces, of velocities $v_\textrm{I}$ and $v_\textrm{II}$, and three media, of permittivities $\epsilon_1$, $\epsilon_2$ and $\epsilon_3$. (b)~Interconnections between the conventional field regions and the generalized, hybrid-field regions around the physical discontinuity.}\label{fig:Gen_FDTD_approach}
\end{figure}

\pati{Local Treatment Solution}
We introduce here an alternative scheme that overcomes all of these limitations by performing a \emph{local treatment} of the space-time discontinuities involved in moving structures. The idea is simply to break the uniformity of the generalized Yee-cell scheme in~\cite{deck2022yeecell} by applying it \emph{only around the positions of the space-time discontinuities} while using the conventional Yee scheme (its $\B{v}=0$ particular) everywhere else. This is globally illustrated in Fig.~\ref{fig:3D_Grid} for the 2+1D problem of oblique light incidence on a contra-directionally moving interface, while Fig.~\ref{fig:Gen_FDTD_approach} provides related details using $zt$-plane (1+1D) projection of the 2+1D problem for a more general moving structure involving two interfaces having arbitrary and hence generally different velocities, with Fig.~\ref{fig:Gen_FDTD_approach}(a) showing the global two-interface three-media problem and Fig.~\ref{fig:Gen_FDTD_approach}(b) depicting the interconnections between the conventional field regions and the generalized, hybrid-field regions, corresponding to the staircase bands around the physical discontinuities, in terms of electromagnetic field samples around the interfaces. Obviously, such a treatment provides a straightforward approach to model moving structures with arbitrary multiple or/and varying velocities, as for instance the double-interface structure in Fig.~\ref{fig:Gen_FDTD_approach} and as will be further illustrated in the examples of Sec.~\ref{sec:Validation}.

\pati{Applicability and Limitations}
The applicability  of the ``localization'' represented in Figs.~\ref{fig:3D_Grid} and~\ref{fig:Gen_FDTD_approach} is a priori not trivial. How can one be sure that spurious numerical scattering will not occur at the numerical, stair-case interfaces between the two types of Yee cells [Fig.~\ref{fig:Gen_FDTD_approach}(b)]? It turns out that, as shown in Appendix~\ref{sec:App_Matching} [Eq.~\eqref{eq:Match_Impedance}], such a scattering issue does fortunately not occur, because the generalized Yee cell is inherently matched to its conventional counterpart. The only limitation of the proposed scheme is that it does not allow to model scattering at space-time point singularities between more than two media, such as the tip of the triangular region at the bottom of Fig.~\ref{fig:Gen_FDTD_approach}(a). At such locations, the different generalized Yee cells would indeed overlap and clash. However, that is not a major problem insofar as the space-time resolution of the simulation can always be increased to get as close as desired to the tip of the structure.

\pati{Additional Benefits}
The proposed localized generalized Yee-cell scheme does not only extend the uniform scheme in~\cite{deck2022yeecell} to arbitrary space-time profiles. It also automatically offers, in contrast to the uniform scheme, immediate applicability of standard absorbing conditions, since the edges of the computational domain, assumed to not include space-time discontinuities, are then associated with the conventional Yee cell, as well as immediate applicability of standard medium dispersion modeling, since the fields in the bulk regions of the different regions are now the real, physical electromagnetic fields\footnote{The forthcoming update equations (Sec.~\ref{sec:Disc_Eq}) do \emph{not} include dispersion because that would make the paper excessively long without bringing substantial benefits: standard FDTD methods to model dispersion, such as the Piecewise-Linear Recursive Convolution (PLRC) and Auxiliary Differential Equation (ADE) methods, may be straightforwardly applied in the proposed scheme.}.

\section{Update Equations}\label{sec:Disc_Eq}
\pati{Discretization of the Generalized Maxwell's Equations}
The final task is to establish the update equations corresponding to the local FDTD scheme elaborated in the previous section, specifically the generalized, hybrid-field update equations to be used in the discontinuity transition regions (the two staircase bands in the case Fig.~\ref{fig:Gen_FDTD_approach}). This implies the 2+1D discretization of the generalized, hybrid-field Maxwell's equations~\eqref{eq:Maxwell_Aux} and constitutive relations~\eqref{eq:Const_Aux}, which gives, under the assumption of s-polarization, 
\begin{subequations}\label{eq:discretize}
\begin{equation}\label{eq:Disc_By}
\begin{split}
B_y|^{n}_{k+\frac{1}{2},i+\frac{1}{2}}=&B_y|^{n-1}_{k+\frac{1}{2},i+\frac{1}{2}}
-\frac{\Delta t}{\Delta z}\left(E^*_x|^{n-\frac{1}{2}}_{k+1,i}-E^*_x|^{n-\frac{1}{2}}_{k,i}\right)
\\&-v\Delta t\left.\frac{\D B_y}{\D z}\right|_{k+\frac{1}{2},i+\frac{1}{2}}^{n-\frac{1}{2}},
\end{split}
\end{equation}
\begin{equation}\label{eq:Disc_Hy}   
H^*_y|_{k+\frac{1}{2},i+\frac{1}{2}}^{n}=\frac{B_y|_{k+\frac{1}{2},i+\frac{1}{2}}^{n}}{\mu|_{k+\frac{1}{2},i+\frac{1}{2}}^{n}}-vD_x|_{k+\frac{1}{2},i+\frac{1}{2}}^{n},
\end{equation}
\begin{equation}\label{eq:Disc_Bz}
\begin{split}
B_z|^{n}_{k+\frac{1}{2},i+\frac{1}{2}}&=B_z|^{n-1}_{k+\frac{1}{2},i+\frac{1}{2}}
+\frac{\Delta t}{\Delta y}\left(E^*_x|^{n-\frac{1}{2}}_{k,i+1}-E^*_x|^{n-\frac{1}{2}}_{k,i}\right)\\&+v\Delta t\left.\frac{\D B_y}{\D y}\right|_{k+\frac{1}{2},i+\frac{1}{2}}^{n-\frac{1}{2}},
\end{split}
\end{equation}
\begin{equation}\label{eq:Disc_Hz}
H^*_z|_{k+\frac{1}{2},i+\frac{1}{2}}^{n}=\frac{B_z|_{k+\frac{1}{2},i+\frac{1}{2}}^{n}}{\mu|_{k+\frac{1}{2},i+\frac{1}{2}}^{n}},
\end{equation}
\begin{equation}\label{eq:Disc_Dx}
\begin{split}
 D_x|^{n+\frac{1}{2}}_{k,i}&=D_x|^{n-\frac{1}{2}}_{k,i}
 +\frac{\Delta t}{\Delta y}\left(H^*_z|^{n}_{k+\frac{1}{2},i+\frac{1}{2}}-H^*_z|^{n}_{k+\frac{1}{2},i-\frac{1}{2}}\right)\\&
-\frac{\Delta t}{\Delta z}\left(H^*_y|^{n}_{k+\frac{1}{2},i+\frac{1}{2}}-H^*_y|^{n}_{k-\frac{1}{2},i+\frac{1}{2}}\right)-v\Delta t \left.\frac{\D D_x}{\D z}\right|_{k,i}^{n}
\end{split}
\end{equation}
and
\begin{equation}\label{eq:Disc_Hx}
E^*_x|_{k,i}^{n+\frac{1}{2}}=\frac{D_x|_{k,i}^{n+\frac{1}{2}}}{\epsilon|_{k.i}^{n+\frac{1}{2}}}-vB_y|_{k,i}^{n+\frac{1}{2}},
\end{equation}
\end{subequations}
where the hybrid (starred) fields~\eqref{eq:star} are discretized as
\begin{subequations}\label{eq:Transition_Star}
    \begin{equation}\label{eq:Transition_Star_Ex}
        E^*_x|_{k,i}^{n+\frac{1}{2}}=E_x|_{k,i}^{n+\frac{1}{2}}-v\frac{\left(B_y|^{n-1}_{k-\frac{1}{2},i+\frac{1}{2}}+B_y|^{n-1}_{k+\frac{1}{2},i+\frac{1}{2}}\right)}{2},
    \end{equation}
    \begin{equation}\label{eq:Transition_Star_Hy}
        H^*_y|_{k+\frac{1}{2},i+\frac{1}{2}}^{n}=H_y|_{k+\frac{1}{2},i+\frac{1}{2}}^{n}-v\frac{\left(D_x|_{k,i}^{n+\frac{1}{2}}+D_x|_{k+1,i}^{n+\frac{1}{2}}\right)}{2}
    \end{equation}
    and
    \begin{equation}\label{eq:Transition_Star_Hz}
        H^*_z|_{k+\frac{1}{2},i+\frac{1}{2}}^{n}=H_z|_{k+\frac{1}{2},i+\frac{1}{2}}^{n}.
    \end{equation}
\end{subequations}

\pati{$v$-Dependent Part Discretization for Stability}
In Eqs.~\eqref{eq:discretize}, we have explicitly discretized only the velocity-independent ($v=0$) parts of Eqs.~\eqref{eq:Maxwell_Aux} and~\eqref{eq:Const_Aux}\footnote{The corresponding (discretized) parts of Eqs.~\eqref{eq:discretize} naturally reduce to the conventional update equations~\eqref{eq:Conv_FDTD} upon explicitly setting $v$ to zero in Eqs.~\eqref{eq:Transition_Star}.}. The discretization of the velocity-dependent ($v\neq 0$) parts must be done by establishing stencils that provide numerical stability. We have found empirically, following~\cite{deck2022yeecell}, that appropriate generalized difference and average update equations are
\begin{subequations}\label{eq:Advection}
\begin{equation}\label{eq:Advection_pB}
    \begin{split}
        \left.\frac{\D B_y}{\D z}\right|_{k+\frac{1}{2},i+\frac{1}{2}}^{n-\frac{1}{2}}&=\frac{1+\mathrm{sgn}(v)}{2}\frac{B_y|^{n-1}_{k+\frac{1}{2},i+\frac{1}{2}}-B_y|^{n-1}_{k-\frac{1}{2},i+\frac{1}{2}}}{\Delta z}\\&+\frac{1-\mathrm{sgn}(v)}{2}\frac{B_y|^{n-1}_{k+\frac{3}{2},i+\frac{1}{2}}-B_y|^{n-1}_{k+\frac{1}{2},i+\frac{1}{2}}}{\Delta z},
    \end{split}
\end{equation}
\begin{equation}\label{eq:Advection_D}
    \begin{split}
    D_x|_{k+\frac{1}{2},i+\frac{1}{2}}^{n}&=\frac{1+\mathrm{sgn}(v)}{2}\frac{D_x|_{k+1,i}^{n-\frac{1}{2}}+D_x|_{k,i}^{n-\frac{1}{2}}}{2}\\&+\frac{1-\mathrm{sgn}(v)}{2}\frac{D_x|_{k,i}^{n-\frac{1}{2}}+D_x|_{k-1,i}^{n-\frac{1}{2}}}{2},
    \end{split}
\end{equation}
\begin{equation}\label{eq:Advection_pBy}
    \begin{split}
        \left.\frac{\D B_y}{\D y}\right|_{k+\frac{1}{2},i+\frac{1}{2}}^{n-\frac{1}{2}}&=\frac{1+\mathrm{sgn}(v)}{2}\frac{B_y|^{n}_{k+\frac{1}{2},i+\frac{1}{2}}-B_y|^{n}_{k+\frac{1}{2},i-\frac{1}{2}}}{\Delta y}\\&+\frac{1-\mathrm{sgn}(v)}{2}\frac{B_y|^{n}_{k+\frac{1}{2},i+\frac{3}{2}}-B_y|^{n}_{k+\frac{1}{2},i+\frac{1}{2}}}{\Delta y},
    \end{split}
\end{equation}
\begin{equation}\label{eq:Advection_pD}
    \begin{split}
    \left.\frac{\D D_x}{\D z}\right|_{k,i}^{n}&=\frac{1+\mathrm{sgn}(v)}{2}\frac{D_x|_{k,i}^{n-\frac{1}{2}}-D_x|_{k-1,i}^{n-\frac{1}{2}}}{\Delta z}\\&+\frac{1-\mathrm{sgn}(v)}{2}\frac{D_x|_{k+1,i}^{n-\frac{1}{2}}-D_x|_{k,i}^{n-\frac{1}{2}}}{\Delta z}
    \end{split}
\end{equation}
and
\begin{equation}\label{eq:Advection_B}
    \begin{split}
    B_y|_{k,i}^{n+\frac{1}{2}}&=\frac{1+\mathrm{sgn}(v)}{2}\frac{B_y|_{k-\frac{3}{2},i-\frac{1}{2}}^n+B_y|_{k-\frac{1}{2},i-\frac{1}{2}}^n}{2}\\&+\frac{1-\mathrm{sgn}(v)}{2}\frac{B_y|_{k+\frac{1}{2},i-\frac{1}{2}}^n+B_y|_{k+\frac{3}{2},i-\frac{1}{2}}^n}{2},
    \end{split}
\end{equation}
\end{subequations}
where the \textrm{signum} function conveniently toggles between the first-row ($v>0$) and second-row ($v<0$) expressions for each equation\footnote{Also note that although they have been checked empirically -- No general synthesis technique for determining stable stencils in the discretization of given partial differential equations seems to exist! -- Eqs.~\eqref{eq:Advection} are based on stencils involving differences that are shifted to the direction of motion (sign of $v$), which makes intuitive sense considering ``numerical advection''.}. According to Eqs.~\eqref{eq:Advection}, the width of the generalized Yee cell region(s) varies between four cells or five points and five cells or six points, as illustrated in Fig.~\ref{fig:Gen_FDTD_approach}. That width is determined by the largest stencil in Eqs.~\eqref{eq:Advection}, namely the stencil of Eq.~\eqref{eq:Advection_B}, which is of five points (the points $k-3/2$, $k-1/2$, $k$, $k+1/2$, $k+3/2$), or four cells, or six points or five cells at the half-integer time steps corresponding to a temporal variation (of one cell) of the discretized interface (see Fig.~\ref{fig:Gen_FDTD_approach}). Those width numbers correspond to the global form of Eqs.~\eqref{eq:Advection}, before the assignation of the velocity; once the velocity has been assigned, the effective width is reduced to numbers corresponding to only one side of the numerical interface, i.e., only one side of the transition regions in Fig.~\ref{fig:Gen_FDTD_approach}, since the \textrm{signum} toggles in Eqs.~\eqref{eq:Advection} keep only one of the two rows in the right-hand side terms the equations\footnote{Thus, in the case Fig.~\ref{fig:Gen_FDTD_approach}, the transition region for the negative-velocity interface (zone A) may be reduced to the region at the right of the numerical interface, while the transition region for the positive-velocity interface (zone A) may be reduced to the region at the left of the numerical interface.}.

\begin{figure}[h!] 
    \centering
    \includegraphics[width=\columnwidth]{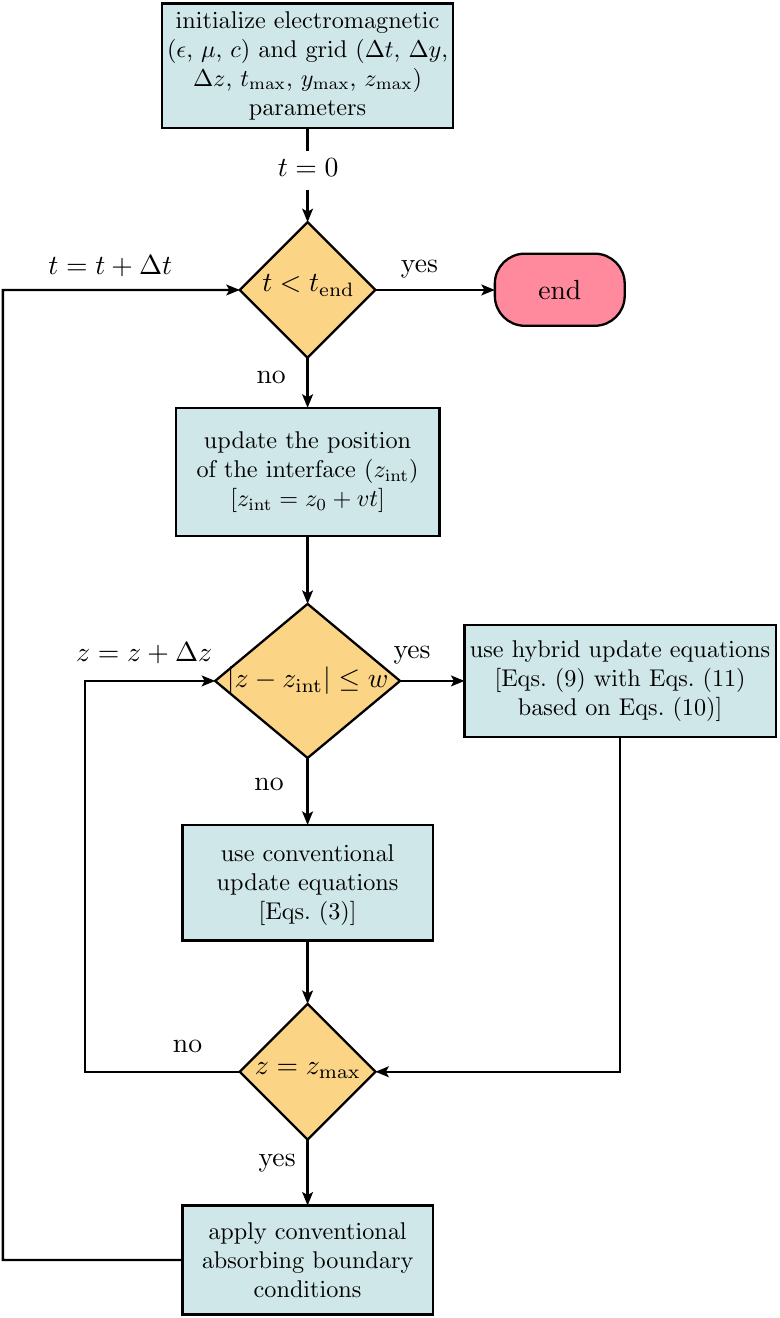}
    \caption{Flow chart describing the equation updating sequence in the proposed scheme. The parameter $w$ corresponds to the width of the transition region (see Fig.~\ref{fig:3D_Grid}).}\label{fig:flowchart}
\end{figure}

\pati{Flowchart}
Figure~\ref{fig:flowchart} shows the update sequence of the field values in the proposed scheme. After initializing the electromagnetic ($\epsilon$, $\mu$ and $c$) and grid ($\Delta t$, $\Delta y$, $\Delta z$, $t_\mt{max}$, $y_\mt{max}$ and $z_\mt{max}$) parameters, the scheme uses the conventional FDTD algorithm~\cite{taflove2005computational} for the static regions and the hybrid field equation algorithm for the interface transition regions (see Figs.~\ref{fig:3D_Grid} and~\ref{fig:Gen_FDTD_approach}).

\section{Stability Analysis}\label{sec:stability_anal}

We now perform a stability analysis of the proposed scheme, using the von Neumann method~\cite{vonneumann1949}. Inserting the test plane wave $\Psi=\Psi_0\zeta\xi$ ($\Psi_0=E_0,B_0,D_0,H_0$), where $\zeta=e^{\alpha\Delta t}$ and $\xi=e^{i k_z\Delta z}=e^{i\theta_z}$, into the update equations~\eqref{eq:discretize}, leads to a $4\times{4}$ matrix system whose determinant's nullification yields
the characteristic equation\footnote{For the sake of conciseness, the following analysis is restricted to the one-dimensional case (wave and modulation propagation in the $z$-direction, with $H_z=B_z=0$). The extension to the (more constraining) two-dimensional and three-dimensional cases is straightforward, although fairly tedious. However, in the case of this paper, where the modulation propagation is restricted to the $z$-direction, the stencil is modified only in that direction, so that $\Delta{z}$ is the most constraining spatial step (more than $\Delta{x}$ and $\Delta{y}$), requiring only appropriate extra dimensional division (by $\sqrt{2}$ or $\sqrt{3}$)~\cite{taflove2005computational}.}~\cite{deck2022yeecell}, 
\begin{subequations}\label{eq:stability_char_eq}
\begin{equation}
    \zeta^2-2b\zeta+d=0,
\end{equation}
where
\begin{equation}
\begin{split}
b=4S^2-n^2(4-3S\beta)-4S(S+n^2\beta)\cos\theta_z\\-n^2S\beta[\cos2\theta_z+2i\sin\theta_z-i\sin2\theta_z]
\end{split}
\end{equation} 
and 
\begin{equation}
    d=e^{-i\theta_z}(1-S\beta+S\beta\cos\theta_z)(S\beta+(1-S\beta)\cos\theta_z+i\sin\theta_z).
\end{equation}
with
\begin{equation}\label{eq:Courant_fact}
    S=c\Delta t/\Delta z
\end{equation}
being $S$ being the Courant factor.
\end{subequations}
\\
\indent Eq.~\eqref{eq:stability_char_eq} is also quadratic in $S$ and may therefore also be written as
\begin{subequations}\label{eq:stability_S}
\begin{equation}
    S^2+pS+q=0,
\end{equation}
where 
\begin{equation}
    p=\frac{(1-e^{i\theta_z})(1-\zeta)n^2\beta}{4\zeta e^{2i\theta_z}+(1-e^{i\theta_z})^2n^2\beta^2}
\end{equation}
and
\begin{equation}
    q=-\frac{e^{i\theta_z}(1-\zeta)^2n^2
    }{\zeta(1-e^{i\theta_z})^2+n^2\beta^2(1-2\cos\theta_z+\cos\theta_z^2)}.
\end{equation}
\end{subequations}
\indent The stability limit corresponds to the magnitude of the attenuation or amplification factor $\zeta$ being equal to one ($|\zeta|=1$) or, equivalently, $\zeta=\tx{e}^{i\phi}$. Inserting  $\zeta=\tx{e}^{i\phi}$ into Eq.~\eqref{eq:stability_S} and solving for $S$, we obtain a solution for $S(\theta_z, \phi, n,\beta)$. The stability limit occurs for $\Psi=\Psi_0$, i.e., when there is no phase accumulation in the test wave. This corresponds to the twofold condition $\zeta=-1$ and $\theta_z=\pi$, which is the same as that for the stationary case~\cite{vonneumann1949}. Enforcing this condition in Eq.~\eqref{eq:stability_S} and solving for $S$ leads to the limit
\begin{equation}\label{eq:S}
    S_\text{max}=\frac{n}{1+n|\beta|},
\end{equation}
with $n$ being the refractive index of the medium and $\beta=v/c$ being the normalized velocity, which we empirically verified by running the scheme for different values of $S$. Equation~\eqref{eq:S} is a \emph{dynamic generalization} of the Courant stability limit, reducing in the stationary limit ($\beta=0$) to the usual result $S_\text{max}=n$, i.e., $\Delta t/\Delta z=n/c$ using Eq.~\eqref{eq:Courant_fact}. In the dynamic regime, we have $S_\text{max}<n$, i.e., $\Delta t/\Delta z<n/c$, which means that stability is more constraining, as might have been expected from the \emph{active} nature of the medium.

\begin{figure}[h!] 
    \centering
    \includegraphics[width=8.6cm]{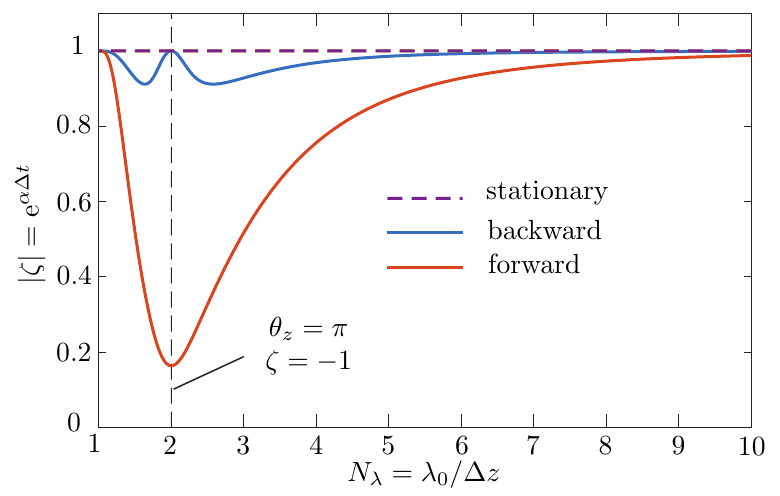}
    \caption{Attenuation factor ($\zeta$) versus number of cells per wavelength ($N_\lambda$) for the stationary case ($\beta=0$) and forward and backward waves (for $\beta>0$) in the dynamics case~[Eq.~\eqref{eq:stability_char_eq}] and $S=S_\tx{max}$, with $n=1.5$ and $\beta=0.3$.}\label{fig:Stability}
\end{figure}

Finally, Fig.~\ref{fig:Stability} plots the wave attenuation at the stability limit $S=S_\tx{max}$. In the stationary regime, $|\zeta|=1$ for any number of cells per wavelength, $N_\lambda=2\pi/(k_z\Delta_z)=\lambda/\Delta z$. In contrast, in the dynamic regime, our generalized scheme exhibits substantial attenuation for $N_\lambda<10$. This numerical (unphysical) attenuation effect is not a problem for a sufficiently fine mesh, $N_\lambda>20$~\cite{deck2022yeecell}.

\section{Validation and Illustrative Examples}\label{sec:Validation}

\pati{Contents}
We validate and illustrate here the proposed scheme via four examples of moving-perturbation structures: a space-time interface under oblique incidence, a space-time wedge, a space-time accelerated interface and a curved space-time interface. The first three examples correspond to canonical structures, from which other structures with arbitrary space-time configurations (space-time uniform and nonuniform slabs, stacks, crystals, gradients, edges, prisms, lenses, complex media with arbitrary geometries) can be formed and which admit exact (closed-form) solutions, given in the appendices, that provide ideal bench-marking for validating the proposed scheme. The fourth example deals with a more involved structure that admits no exact solution and illustrates the capability of the scheme to handle arbitrarily complex structures.

\subsection{Space-Time Interface}
\begin{figure}[h!] 
    \centering
    \includegraphics[width=8.6cm]{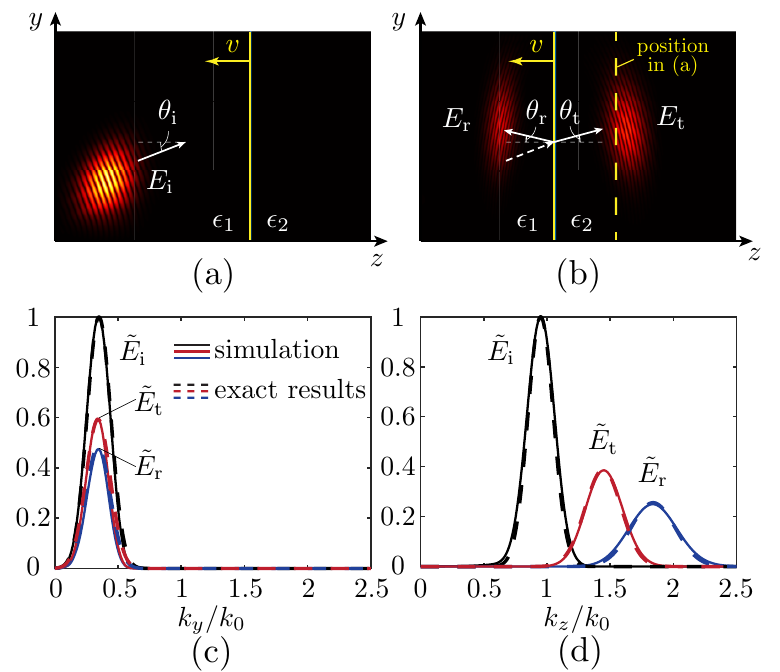}
    \caption{Scattering of the obliquely incident modulated Gaussian pulse \mbox{$E_\mathrm{i}=e^{-i\omega_\mt{i}t+ik_{\mt{i},y}y+ik_{\mt{i},z}z}e^{-(t/\tau_\mathrm{i})^2-(y/\sigma_{\mt{i},y})^2-(z/\sigma_{\mt{i},z})^2}$}, where $k_{\mt{i},y}=k_\mt{i}\sin\theta_\mt{i}$, $k_{\mt{i},z}=k_\mt{i}\cos\theta_\mt{i}$, $\tau_\mt{i}=5T_\mt{i}$ and $\theta_\mt{i}=20^{\circ}$, at a single interface moving at the uniform velocity $v=-0.3c$ between two media with $\epsilon_1=1$, $\epsilon_2=3$ and $\mu=1$, with the FDTD parameters $\Delta z=\Delta y=\lambda_0/20$ and $\Delta t=\Delta z/(5c)$. Snapshots of (a)~the incident and (b)~scattered waves in the spatial domain, corresponding to the reflection and transmission angles $\theta_\mt{r}=10.85^{\circ}$ and $\theta_\mt{t}=13.42^{\circ}$. Fourier transforms of the incident and scattered waves along (c)~the $y$- and (d)~$z$-directions.}\label{fig:Single_2d_Sub}
\end{figure}

Figure~\ref{fig:Single_2d_Sub} presents the results for the space-time interface, under oblique incidence (scenario represented in Fig.~\ref{fig:3D_Grid}). Figures~\ref{fig:Single_2d_Sub}(a) and~(b) qualitatively show the scattering of a space-time pulse into reflected and transmitted parts, while Figs.~\ref{fig:Single_2d_Sub}(c) and~(d) provide corresponding quantitative spectral information. All the simulation results perfectly agree with the exact results, given by Eq.~\eqref{eq:Pulse_Scat_inv}\footnote{The scattering angles ($\theta_\mt{r,t}$) may be found by inserting the corresponding $k$-values ($k_{\mt{r,t};y,z}$) taken at the maxima of the spectral pulses in Figs.~\ref{fig:Single_2d_Sub}(c) and~(d) into Eqs.~\eqref{eq:k_Transition}, and isolating the desired angle.}. Note the spatial compression of the scattered pulses in Fig.~\ref{fig:Single_2d_Sub}(b) and corresponding ($\B{k}$) spectral expansion, which are due to contra-directional Doppler scattering for the reflected pulse and increasing-permittivity contrast for the transmitted pulse.

\subsection{Space-Time Wedge}\label{sec:ST_wedge}
\begin{figure}[h!]
    \centering
    \includegraphics[width=6cm]{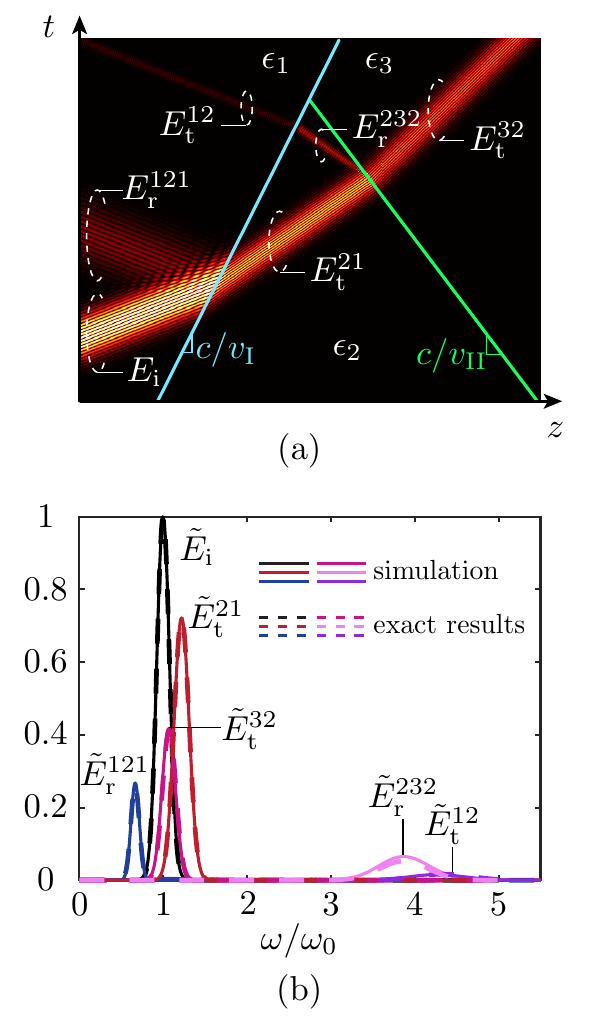}
    \caption{Scattering of the modulated Gaussian pulse \mbox{$E_\mt{i}=e^{-i\omega_\mt{i}t+ik_\mt{i}z}e^{-(t/\tau_\mt{i})^2}e^{-(z/\sigma_\mt{i})^2}$}(normal incidence), with $\tau_\mt{i}=3T_\mt{i}$, at two interfaces of different velocities, $v_1=0.2c$ and $v_2=-0.3c$, separating three media with $\epsilon_1=1$, $\epsilon_2=3$, $\epsilon_2=6$ and $\mu=1$ with the FDTD parameters of $\Delta z=\lambda_0/85$ and $\Delta t=\Delta z/(5c)$ in (a)~space-time domain and in (b)~frequency domain.}\label{fig:Wedge}
\end{figure}

Figure~\ref{fig:Wedge} presents the results for the space-time wedge, which consists of two interfaces of different velocities, meeting at a specific point of spacetime, and which is assumed to be normally illuminated. Figure~\ref{fig:Wedge}(a) qualitatively shows the multiple scattering of a pulse in space and time\footnote{While Figs.~\ref{fig:Single_2d_Sub}(a) and~(b) use a 2D space representation to emphasize the obliqueness of scattering, this figure, pertaining to normal incidence, uses a 1+1D space-time representation to emphasize multiple scattering in space-time.} on the wedge, while Fig.~\ref{fig:Wedge}(b) provides corresponding quantitative spectral information. Again, all the simulation results perfectly agree with the exact results, given by a successive application of Eq.~\eqref{eq:Pulse_Scat_inv} at each of the two interfaces. Note that the pulse width within the wedge progressively decreases at each reflection event, as expected from the contra-moving nature of the corresponding scattering.

\subsection{Space-Time Accelerated Interface}\label{sec:acc_interf}
\begin{figure}[h!]
    \centering
    \includegraphics[width=6cm]{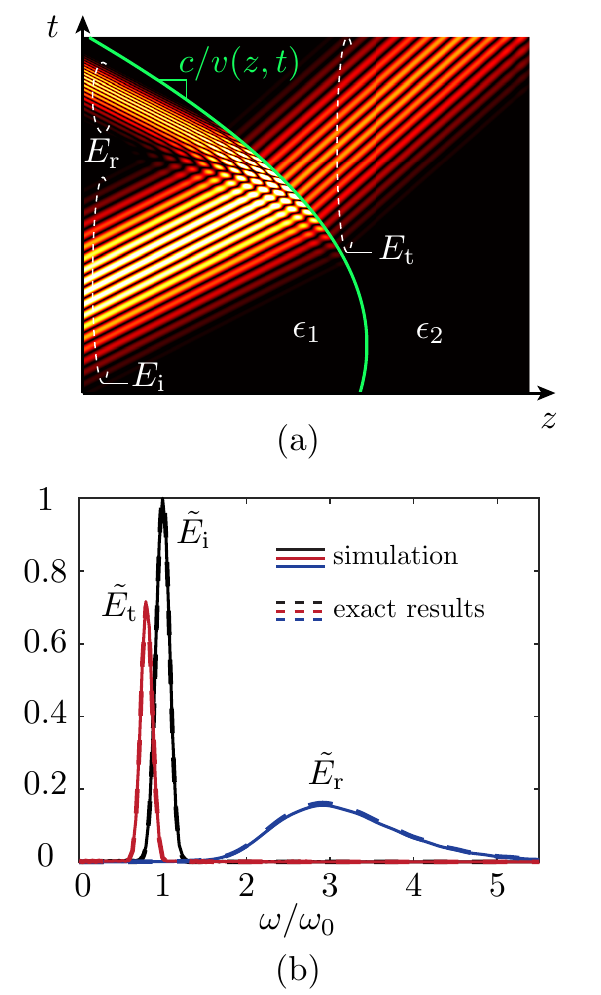}
    \caption{Scattering of the modulated Gaussian pulse \mbox{$E_\mt{i}=e^{-i\omega_\mt{i}t+ik_\mt{i}z}e^{-(t/\tau_\mt{i})^2}e^{-(z/\sigma_\mt{i})^2}$} (normal incidence), with $\tau_\mt{i}=3T_\mt{i}$, at a single interface moving with the constant proper acceleration $a'=-0.2c$ between two media with $\epsilon_1=1$, $\epsilon_2=3$ and $\mu=1$, with the FDTD parameters of $\Delta z=\lambda_0/100$ and $\Delta t=\Delta z/(5c)$. (a)~Space-time evolution of the pulse. (b)~ Fourier transforms of the scattered and the incident pulses.}\label{fig:Acc}
\end{figure}

Figure~\ref{fig:Acc} presents the results for the space-time accelerated interface, again under normal incidence. Figure~\ref{fig:Acc}(a) qualitatively shows the scattering of a pulse in the space-time diagram, while Fig.~\ref{fig:Acc}(b) provides corresponding quantitative spectral information. As in the previous two examples, all the simulation results perfectly agree with the exact results, given this time by Eqs.~\eqref{eq:ASTEM_Scattered}, with the results in Fig.~\ref{fig:Acc}(b) being obtained by numerical Fourier transformation of these equations. Note the space-time chirping effect induced by the nonuniformity (acceleration) of the interface, which is apparent in the variation of the separation between the trajectory crests (or troughs) of the scattered pulse in time and space\footnote{This chirping effect is much more pronounced in the reflected pulse than in the transmitted pulse. This is because, for the prevailing case contra-directional scattering ($\beta=-|\beta|$), the Doppler frequency shift is much greater in reflection [$\omega_\mt{r}=\omega_\mt{i}(1+n_1|\beta|)/(1-n_1|\beta|)$] than transmission [$\omega_\mt{t}=\omega_\mt{i}(1+n_1|\beta|)/(1+n_2|\beta|)$] . The opposite would be true for co-directional scattering.}.

\subsection{Curved Space-Time Interface}
\begin{figure}[h!] 
    \centering
    \includegraphics[width=\columnwidth]{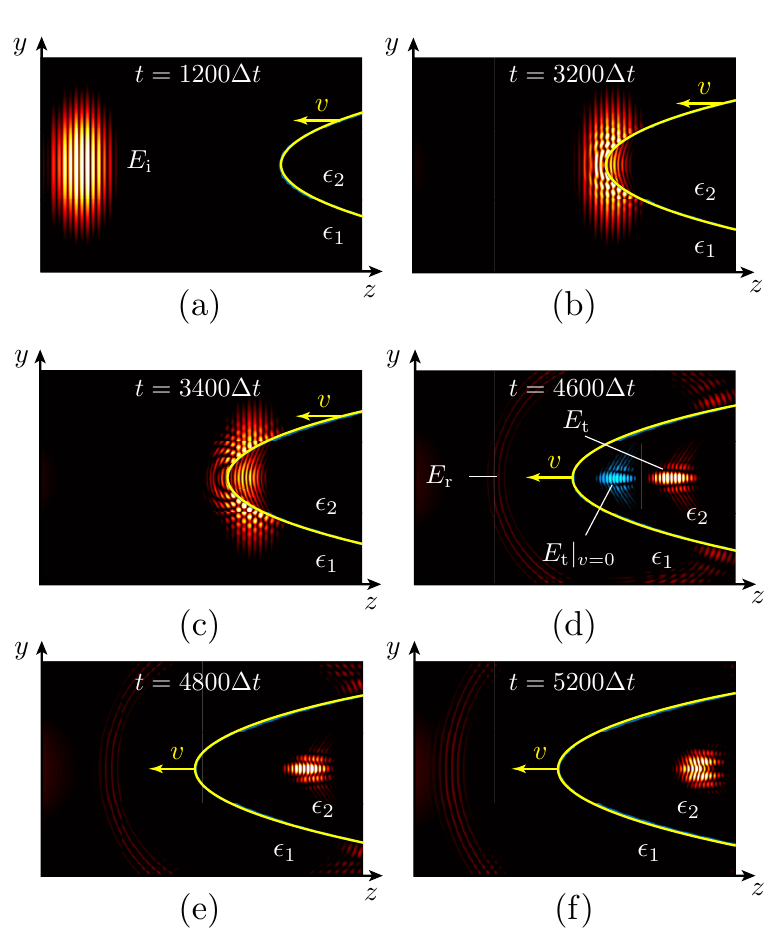}
    \caption{Scattering of the normally incident modulated Gaussian pulse \mbox{$E_\mathrm{i}=e^{-i\omega_\mt{i}t+ik_{\mt{i},z}z}e^{-(t/\tau_\mathrm{i})^2-(y/\sigma_{\mt{i},y})^2}$}, with $\tau_\mt{i}=6T_\mt{i}$, at a single parabolically curved interface between two media, with $\epsilon_1=1$, $\epsilon_2=3$ and $\mu=1$, moving at the uniform velocity $v=-0.3c$, with the FDTD parameters $\Delta z=\Delta y=\lambda_0/20$ and $\Delta t=\Delta z/(5c)$. Snapshots at (a)~$t=1200\Delta t$, (b)~$t=3200\Delta t$, (c)~$t=3400\Delta t$, (d)~$t=4600\Delta t$, (e)~$t=4800\Delta t$ and (f)~$t=5200\Delta t$. The panel~(d) also shows the focal spot corresponding to a stationary ($v=0$) parabolic interface.}\label{fig:Curved_2d}
\end{figure}

Figure~\ref{fig:Curved_2d} presents the results for the curved space-time interface under normal incidence, with Fig.~\ref{fig:Curved_2d}~(a)-(f) corresponding to different successive temporal snapshots. The interface curve could be of arbitrary shape, but we have chosen it here to be parabolic to emphasize physical effects in the forthcoming discussion. Note that the effect of motion is to increase the focal distance [Fig.~\ref{fig:Curved_2d}(d)], as expected from the fact that refracted waves are deflected (or refracted rays are rotated) towards the direction of motion at a moving interface~\cite{Deck_2018_wave}. Another, more subtle effect may be noted. One might have a priori expected some dynamic aberration (stretching of the focal spot), akin to chromatic aberration, from the fact that new frequencies are being created by motion. However, the dynamic focal spot appears to be of very comparable size with the stationary one. This might be explained by the compensating effect of transmitted wavelength compression in the present case of negative  modulation velocity~\cite{caloz2019spacetime2}.
\section{Conclusion}\label{sec:Conclusion}
\pati{Summary} We have presented a generalized FDTD scheme to simulate moving electromagnetic structures with arbitrary space-time configurations and demonstrated the validity of this scheme against exact results for a few moving-perturbation space-time structures.

\pati{Significance} The proposed scheme fills a gap in both the open literature, whose related attempts were restricted to impenetrable, single and uniform velocity moving structures, and in commercial software capabilities, which previously included the simulation of pure-space and pure-time but not space-time structures.

\appendices\label{sec:appendices}

\section{Scattering from Moving Perturbation Interface}\label{sec:App_USTEM}
\pati{Introduction}
We derive here the scattering coefficients or fields and deflection angles pertaining to plane wave and Gaussian pulse oblique incidence on a (uniformly) moving perturbation interface for bench-marking some results in Secs.~\ref{sec:Failure} and~\ref{sec:Validation}. The basic formulas for the plane wave case are already given in~\cite{Kunz_1980_plane}, but we re-derive them here for the sake of completeness and notational convenience, along with new formulas for the pulse wave case. 

\pati{Forms of the Fields}
Under the paper's assumption of s-polarization and for the coordinate system selected in the paper (e.g., Fig.~\ref{fig:Single_2d_Sub}), the incident (i), reflected (r) and transmitted (t) electric and magnetic fields have the forms
\begin{subequations}\label{eq:Uni_ass_E}
\begin{equation}\label{eq:Uni_ass_Ei} 
    \B{E_\mt{i}}=\hat{x}A_\mt{i}e^{i\phi_\mt{i}},
\end{equation}
\begin{equation}\label{eq:Uni_ass_Er}
    \B{E_\mt{r}}=\hat{x}A_\mt{r}e^{i\phi_\mt{r}},
\end{equation}
\begin{equation}\label{eq:Uni_ass_Et}
    \B{E_\mt{t}}=\hat{x}A_\mt{t}e^{i\phi_\mt{t}} ,
\end{equation}
\end{subequations}
and
\begin{subequations}\label{eq:Uni_ass_H}
\begin{equation}\label{eq:Uni_ass_Hi}
    \B{H_\mt{i}}=\left(\hat{y}\cos\theta_\mt{i}-\hat{z}\sin\theta_\mt{i}\right)\frac{A_\mt{i}}{\eta_1}e^{i\phi_\mt{i}},
\end{equation}
\begin{equation}\label{eq:Uni_ass_Hr}
    \B{H_\mt{r}}=\left(-\hat{y}\cos\theta_\mt{r}-\hat{z}\sin\theta_\mt{r}\right)\frac{A_\mt{r}}{\eta_1}e^{i\phi_\mt{r}},
\end{equation}
\begin{equation}\label{eq:Uni_ass_Ht}
    \B{H_\mt{t}}=\left(\hat{y}\cos\theta_\mt{t}-\hat{z}\sin\theta_\mt{t}\right)\frac{A_\mt{t}}{\eta_2}e^{i\phi_\mt{t}},
\end{equation}
\end{subequations}
where $\eta_1$ and $\eta_2$ are the impedances of the incidence/reflection and transmission media, respectively, $\theta_\mt{i,r,t}$ are the angles with respect to the $z$ axis, and where the space-time dependent phases are
\begin{subequations}\label{eq:Uni_phi}
\begin{equation}\label{eq:Uni_phii}
    \phi_\mt{i}=k_{\mt{i},y} y+k_{\mt{i},z} z-\omega_\mt{i}t,
\end{equation}
\begin{equation}\label{eq:Uni_phir}
    \phi_\mt{r}=k_{\mt{r},y} y-k_{\mt{r},z} z-\omega_\mt{r}t,
\end{equation}
\begin{equation}\label{eq:Uni_phit}
    \phi_\mt{t}=k_{\mt{t},y} y+k_{\mt{t},z} z-\omega_\mt{t}t.
\end{equation}
\end{subequations}

\pati{Comoving Frame Expressions}
The problem is greatly simplified by applying the frame hopping technique~\cite{van2012relativity}, i.e., by transforming the electric and magnetic fields into corresponding expressions in the comoving frame, where the interface is stationary, applying there the usual (stationary) boundary conditions, and inverse-Lorentz transforming the resulting fields. Lorentz-transforming the fields in Eqs.~\eqref{eq:Uni_ass_E} and~\eqref{eq:Uni_ass_H} yield~\cite{van2012relativity,Kunz_1980_plane}
\begin{subequations}\label{eq:Uni_K'}
\begin{equation}\label{eq:Uni_Ei'}
    \B{E'_\mt{i}}=\hat{x}\gamma A_\mt{i}(1-n_1\beta\cos\theta_\mt{i})e^{i\phi'_\mt{i}},
\end{equation}
\begin{equation}\label{eq:Uni_Er'}
    \B{E'_\mt{r}}=\hat{x}\gamma A_\mt{r}(1+n_1\beta\cos\theta_\mt{r})e^{i\phi'_\mt{r}},
\end{equation}
\begin{equation}\label{eq:Uni_Et'}
    \B{E'_\mt{t}}=\hat{x}\gamma A_\mt{t}(1-n_2\beta\cos\theta_\mt{t})e^{i\phi'_\mt{t}},
\end{equation}
and
\begin{equation}\label{eq:Uni_Hi'}
    \B{H'_\mt{i}}=\hat{y}\gamma\frac{A_\mt{i}}{\eta_1}(\cos\theta_\mt{i}-n_1\beta)e^{i\phi'_\mt{i}}-\hat{z}\frac{A_\mt{i}}{\eta_1}\sin\theta_\mt{i}e^{i\phi'_\mt{i}},
\end{equation}
\begin{equation}\label{eq:Uni_Hr'}
    \B{H'_\mt{r}}=-\hat{y}\gamma\frac{A_\mt{r}}{\eta_1}(\cos\theta_\mt{r}+n_1\beta)e^{i\phi'_\mt{r}}-\hat{z}\frac{A_\mt{r}}{\eta_1}\sin\theta_\mt{r}e^{i\phi'_\mt{r}},
\end{equation}
\begin{equation}\label{eq:Uni_Ht'}
    \B{H'_\mt{t}}=\hat{y}\gamma\frac{A_\mt{t}}{\eta_2}(\cos\theta_\mt{t}-n_2\beta)e^{i\phi'_\mt{t}}-\hat{z}\frac{A_\mt{t}}{\eta_2}\sin\theta_\mt{t}e^{i\phi'_\mt{t}},
\end{equation}
\end{subequations}
where $\beta=v/c$ is the normalized velocity of the interface, $\gamma=1/\sqrt{1-\beta^2}$ is the corresponding Lorentz factor, and the primed phase expressions are identical to those in Eqs.~\eqref{eq:Uni_phi} but with primes added everywhere.

\pati{Boundary Conditions, Inverse Transformation and Results}
Applying the stationary boundary conditions at the interface in the comoving frame, namely enforcing there the continuity of the primed tangential components of the electric and magnetic fields, and inverse-Lorentz transforming the results back to the laboratory frame leads to the reflection and transmission (or scattering) coefficients
\begin{subequations}\label{eq:Uni_Scat_Coeff}
    \begin{equation}\label{eq:Uni_Scat_Coeff_Ref}
        \Gamma=\frac{A_\mt{r}}{A_\mt{i}}=a_\mt{r}\frac{Z_2-Z_1}{Z_2+Z_1}
    \end{equation}
    and
    \begin{equation}\label{eq:Uni_Scat_Coeff_Trans}
       T=\frac{A_\mt{r}}{A_\mt{i}}=a_\mt{t}\frac{2Z_2}{Z_2+Z_1},
    \end{equation}
    where
    \begin{equation}\label{eq:ar}
       a_\mt{r}=\frac{1-n_1\beta\cos\theta_\mt{i}}{1+n_1\beta\cos\theta_\mt{r}},
    \end{equation}
    \begin{equation}\label{eq:at}
       a_\mt{t}=\frac{1-n_1\beta\cos\theta_\mt{i}}{1-n_2\beta\cos\theta_\mt{t}},
    \end{equation}
    \begin{equation}\label{eq:Z1}
       Z_1=\frac{1-n_1\beta\cos\theta_\mt{i}}{\cos\theta_\mt{i}-n_1\beta}\eta_1
    \end{equation}
    and
    \begin{equation}\label{eq:Z2}
       Z_2=\frac{1-n_2\beta\cos\theta_\mt{t}}{\cos\theta_\mt{t}-n_2\beta}\eta_2,
    \end{equation}
\end{subequations}
and to the scattered temporal and spatial frequencies

\begin{subequations}\label{eq:Freq_Transition}
    \begin{equation}\label{eq:Freq_Transition_refl}
        \omega_\mt{r}=\frac{1-n_1\beta\cos\theta_\mt{i}}{1+n_1\beta\cos\theta_\mt{r}}\omega_\mt{i},
    \end{equation}
    \begin{equation}\label{eq:Freq_Transition_trans}
        \omega_\mt{t}=\frac{1-n_1\beta\cos\theta_\mt{i}}{1-n_2\beta\cos\theta_\mt{t}}\omega_\mt{i},
    \end{equation}
\end{subequations}
and
\begin{subequations}\label{eq:k_Transition}
\begin{equation}
    k_{\mt{r},z}=\frac{1/n_1-\beta\cos\theta_\mt{i}}{1/n_1+\beta\cos\theta_\mt{r}}k_{\mt{i},z}, \quad k_{\mt{r},y}=k_{\mt{i},y},
\end{equation}
\begin{equation}
    k_{\mt{t},z}=\frac{1/n_1-\beta\cos\theta_\mt{i}}{1/n_2-\beta\cos\theta_\mt{t}}k_{\mt{i},z}, \quad k_{\mt{t},y}=k_{\mt{i},y},
\end{equation}
\end{subequations}
corresponding to reflection and transmission deflection (or scattered) angles
\begin{subequations}\label{eq:Uni_Scat_Angl}
    \begin{equation}\label{eq:Uni_Scat_Angl_Ref}
        \cos\theta_\mt{r}=\frac{(n_1^2\beta^2+1)\cos\theta_\mt{i}-2n_1\beta}{(n_1\beta-\cos\theta_\mt{i})^2+\sin^2\theta_\mt{i}}
    \end{equation}
    and
    \begin{equation}\label{eq:Uni_Scat_Angl_Trans}
    \begin{split}
        \cos\theta_\mt{t}&=\frac{n_1^2\beta\sin^2\theta_\mt{i}}{n_2(n_1\beta-\cos\theta_\mt{i})^2+n_2\sin^2\theta_\mt{i}}+\\&\frac{(1+n_1\beta\cos\theta_\mt{i})\sqrt{n_2^2(n_1\beta-\cos\theta_\mt{i})^2+(n_2^2-n_1^2)\sin^2\theta_\mt{i}}}{n_2(n_1\beta-\cos\theta_\mt{i})^2+n_2\sin^2\theta_\mt{i}}.
    \end{split}
    \end{equation}
\end{subequations}

\pati{Extension to a Pulse}
We next wish to determine how the interface a Gaussian pulsed version of the plane wave in~\eqref{eq:Uni_ass_Ei}, namely
\begin{equation}\label{eq:Pulse_Relation_direct}
        E_\mt{i}(y,z)=\hat{x}A_\mt{i}e^{ik_{\mt{i},y}y+ik_{\mt{i},z}z}e^{-(y/\sigma_y)^2}e^{-(z/\sigma_z)^2}.
 \end{equation}
which is illustrated in Fig.~\ref{fig:Single_2d_Sub}. Since the oscillatory part of this pulse is just the plane wave in~\eqref{eq:Uni_ass_Ei}, the corresponding scattered angles are also given by Eqs.~\eqref{eq:Uni_Scat_Angl}. In contrast, the corresponding scattering field magnitudes are given by
\begin{equation}\label{eq:Pulse_Scat_dir}
   E_\mt{r,t}(y,a_{\mt{r},\mt{t}}z)=\hat{x}\{\check{\Gamma}(z),\check{T}(z)\}*E_\mt{i}(y,z),
\end{equation}
where $\check{\Gamma}(z)$ and $\check{T}(z)$ are impulse responses (the medium is linear time invariant in terms of space at any given time) whose $z$ dependence are related to the arrival and scattering timing of the pulse on the interface. 

The functions $\check{\Gamma}(z)$ and $\check{T}(z)$ are not known a priori. However, Fourier-transforming Eq.~\eqref{eq:Pulse_Scat_dir} yields
\begin{equation}\label{eq:Pulse_Scat_inv}
       \Tilde{E}_\mt{r,t}(k_y,a_{\mt{r},\mt{t}}k_z)=\hat{x}\{\Gamma(k_z),T(k_z)\}\Tilde{E}_\mt{i}(k_y,k_z)/a_{\mt{r},\mt{t}},
\end{equation}
where 
\begin{equation}\label{eq:Pulse_Relation_inverse}
    \Tilde{E}_\mt{i}(k_y,k_z)=\hat{x}\frac{\sigma_y\sigma_z}{16\pi}A_\mt{i}e^{-(\sigma_y(k_y-k_{\mt{i},y}))^2/4}e^{-(\sigma_z(k_z-k_{\mt{i},z}))^2/4}
\end{equation}
is the Fourier transform of Eq.~\eqref{eq:Pulse_Relation_direct} and where the transfer functions $\Gamma(k_z)$ and $T(k_z)$ are given by Eqs.~\eqref{eq:Uni_Scat_Coeff_Ref} with $k_z=k_{\mt{r},z}=n_1 k_0\cos\theta_\mt{r}$ and~\eqref{eq:Uni_Scat_Coeff_Trans} with $k_z=k_{\mt{t},z}=n_2 k_0\cos\theta_\mt{t}$. The scattered field is then obtained by inverse Fourier-transforming this relation.

\section{Natural Enforcement of Pure-Space \\ and Pure-Time Boundary Conditions \\ in Conventional FDTD}\label{sec:App_Natur_BCs}

\pati{Introduction}
We show here that the conventional FDTD scheme~\cite{taflove2005computational} automatically enforces the continuity of the (tangential) $\B{E}$ and $\B{H}$ fields at a pure-space (or stationary) interface, as well-known from routine FDTD simulations, and the continuity of the $\B{D}$ and $\B{B}$ fields at a pure-time (or instantaneous) interface~\cite{Morgenthaler_1958,caloz2019spacetime2}, which has been much less studied and is hence much less known, as claimed in Sec.~\ref{sec:Failure}.

\begin{figure}[h!]
    \centering
    \includegraphics[width=8.6cm]{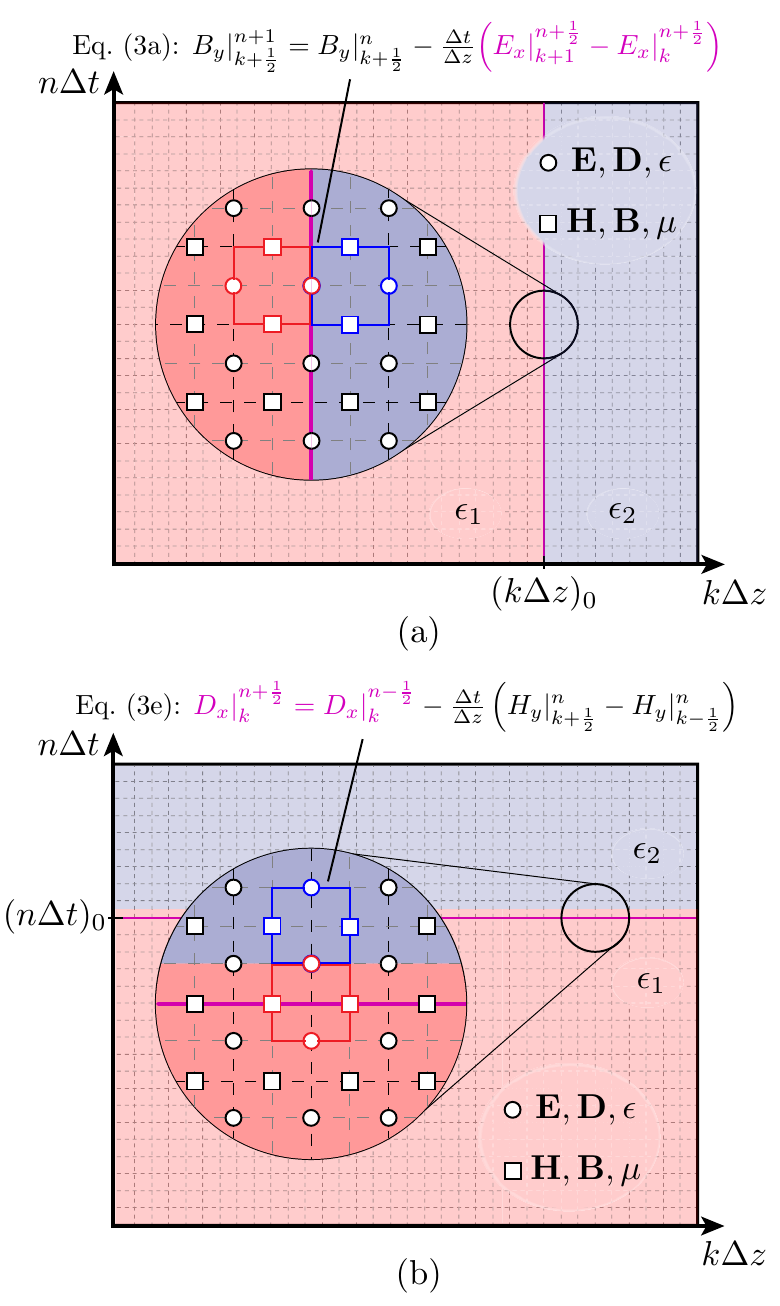}
    \caption{Interface between two different dielectric media (of permittivities $\epsilon_1$ and $\epsilon_2$) in the conventional Yee-cell FDTD grid  for (a)~a pure-space (or stationary) discontinuity, with continuity of the (tangential) $\B{E}$ field, and (b)~a pure-time (or instantaneous) discontinuity, with continuity of the $\B{D}$ field.}\label{fig:App_Yee}
\end{figure}

\pati{Boundary Positioning}
Figure~\ref{fig:App_Yee} shows the two interfaces in the conventional, staggered Yee-cell FDTD computational grid for the case of an interface between two different dielectric media (of permittivities $\epsilon_1$ and $\epsilon_2$), with Figs.~\ref{fig:App_Yee}(a) and~\ref{fig:App_Yee}(b) respectively pertaining to the pure-space and pure-time cases. Notice that in the present case of a dielectric discontinuity, the interfaces are positioned across electric ($\B{E}$,$\B{D}$ and $\epsilon$ space integer and time half-integer) grid points, whereas in the case of a magnetic discontinuity (between media of permeabilities $\mu_1$ and $\mu_2$), they would be positioned across magnetic ($\B{H}$, $\B{B}$ and $\mu$ space half-integer and time integer) grid points.

\pati{Pure-Space Boundary Conditions} 
In the case of the pure-space discontinuity [Fig.~\ref{fig:App_Yee}(a)],  we shall restrict our attention, without loss of generality, to the case of normal incidence ($z$-direction), whose spatial evolution for a dielectric discontinuity involves the fields $E_x$ and $B_y$ according to Eqs.~\eqref{eq:Maxwell_E} and~\eqref{eq:Const_Iso_eps} and corresponds to the update equation~\eqref{eq:Conv_FDTD_By}. The red and blue squares in the inset of the figure highlight the fields involved in that equation at two successive spatial iterations and show that the $E_x$ field sample at the interface is common to the two iterations, which demonstrates the continuity of (tangential) $\B{E}$. An analogous argument with Eq.~\eqref{eq:Conv_FDTD_D} (with $H_z=0$) replacing Eq.~\eqref{eq:Conv_FDTD_By} would demonstrate the continuity of (tangential) $\B{H}$.

\pati{Pure-Time Boundary Conditions}
The case of the pure-time discontinuity [Fig.~\ref{fig:App_Yee}(b)] whose temporal evolution for a dielectric discontinuity involves the fields $\B{D}$ and $\B{H}$ according to Eqs.~\eqref{eq:Maxwell_H} and again~\eqref{eq:Const_Iso_eps} corresponds to the update equation~\eqref{eq:Conv_FDTD_D}. The red and blue squares in the inset of the figure highlight the fields involved in that equation at two successive temporal iterations and show that the $\B{D}$ field sample at the interface is common to the two iterations, which demonstrates the continuity of $\B{D}$. An analogous argument with Eq.~\eqref{eq:Conv_FDTD_By} replacing Eq.~\eqref{eq:Conv_FDTD_D} would demonstrate the continuity of $\B{B}$.

\section{Inherent Matching of the Generalized Yee cell \\ with its Conventional Counterpart}\label{sec:App_Matching}
\pati{Introduction} We show here that, as announced in Sec.~\ref{sec:Local_Treat}, the generalized Yee cell is inherently matched with the conventional Yee cell, which ensures the applicability of the localized scheme.

\pati{Spectral Maxwell's and Constitutive Equations}
This demonstration is most easily done in the spectral domain. Fourier-transforming the generalized, hybrid Maxwell's equations~\eqref{eq:Maxwell_Aux} and corresponding constitutive relations~\eqref{eq:Const_Aux} yields, for s-polarization,
\begin{subequations}\label{eq:Match_Fourier}
    \begin{equation}\label{eq:Match_Fourier_By}
        \omega \Tilde{B}_y=k_z \Tilde{E}^*_x+k_z v\Tilde{B}_y,
    \end{equation}
    \begin{equation}\label{eq:Match_Fourier_Bz}
        \omega \Tilde{B}_z=-k_y \Tilde{E}^*_x-k_y v\Tilde{B}_y,
    \end{equation}
    \begin{equation}\label{eq:Match_Fourier_D}
        \omega \Tilde{D}_x=k_z\Tilde{H}^*_y-k_y\Tilde{H}^*_z+k_z v\Tilde{D}_x,
    \end{equation}
\end{subequations}
and
\begin{subequations}\label{eq:Const_Aux_Expanded}
    \begin{equation}\label{eq:Const_Aux_Expanded_By}
        \Tilde{B}_y=\mu \Tilde{H}^*_y+\mu v \Tilde{D}_x,
    \end{equation}
    \begin{equation}\label{eq:Const_Aux_Expanded_D}
        \Tilde{D}_x=\epsilon \Tilde{E}^*_y+\mu v \Tilde{B}_y.
    \end{equation}
\end{subequations}

\pati{Impedance Derivation}
The related impedance may be found first eliminating $\Tilde{D}_x$ upon inserting Eq.~\eqref{eq:Const_Aux_Expanded_D} into Eq.~\eqref{eq:Match_Fourier_D}, which gives
\begin{equation}\label{eq:Match_D_Rel}
    (\omega-k_z v)(\epsilon \Tilde{E}^*_y+\mu v \Tilde{B}_y)=k_z\Tilde{H}^*_y-k_y\Tilde{H}^*_z,
\end{equation}
and further eliminating $\Tilde{B}_y$ upon substituting $\Tilde{B}_y$ from Eq.~\eqref{eq:Match_Fourier_By} into that relation [Eq.~\eqref{eq:Match_D_Rel}], which leads to
\begin{equation}\label{eq:Match_EHstar}
    \omega \epsilon \Tilde{E}^*_x=k_z \Tilde{H}^*_y-k_y \Tilde{H}^*_z.
\end{equation}
Dividing then both sides of this equation by $E^*_x$ yields then the following expression involving the ratios of the electric field to the magnetic fields:
\begin{equation}\label{eq:Match_EHstar}
    \omega \epsilon =k_z \frac{\Tilde{H}^*_y}{\Tilde{E}^*_x}-k_y \frac{\Tilde{H}^*_z}{\Tilde{E}^*_x}.
\end{equation}
Finally, substituting $k_z=nk_0\cos\theta$, $k_y=nk_0\sin\theta$, $\Tilde{H}^*_y=\Tilde{H}^*_0\cos\theta$ and $\Tilde{H}^*_z=-\Tilde{H}^*_0\sin\theta$, with $\theta$ being the direction of wave propagation in the $yz$-plane, in this equation results into
\begin{equation}\label{eq:Match_Impedance_Simple}
    \omega\epsilon=\left(\frac{\cos^2\theta \Tilde{H}^*_0}{\Tilde{E}^*_x}+\frac{\sin^2\theta \Tilde{H}^*_0}{\Tilde{E}^*_x}\right)nk_0=\frac{nk_0 \Tilde{H}^*_0}{\Tilde{E}^*_x}=\frac{nk_0}{\eta^*}.
\end{equation}
This relation may be rewritten in terms of the impedance as
\begin{equation}\label{eq:Match_Impedance}
    \eta^*=\frac{\Tilde{E}^*_x}{\Tilde{H}^*_0}=\frac{nk_0}{\omega \epsilon}=\frac{\Tilde{E}_x}{\Tilde{H}_0}=\eta,
\end{equation}
revealing that the generalized Yee cell has the same impedance as its conventional counterpart, and is hence impedance-matched to it, which ensure that no spurious numerical scattering occurs in the transition regions [two edges of the staircase band in Fig.~\ref{fig:Gen_FDTD_approach}(b)] between the two types of Yee cells.

\pati{Velocity Matching}
The generalized Yee cell should also be velocity-matched to the conventional Yee cell for it would not the velocity would abruptly change in the transition regions [two edges of the staircase band in Fig.~\ref{fig:Gen_FDTD_approach}(b)], which would induce spurious numerical frequency transitions. Such velocity matching is also true. This is immediately seen upon noting that the $\B{E}^*$ and $\B{H}^*$ fields [Eqs.~\eqref{eq:star}] have the same phase as their physical counterparts, since the $\B{E}$ and $\B{B}$ fields and the $\B{H}$ and $\B{D}$ fields have the same phase and $\B{v}$ is real, and have hence also the same phase velocity (or refractive index).

\begin{figure}
    \centering
    \includegraphics[width=8.6cm]{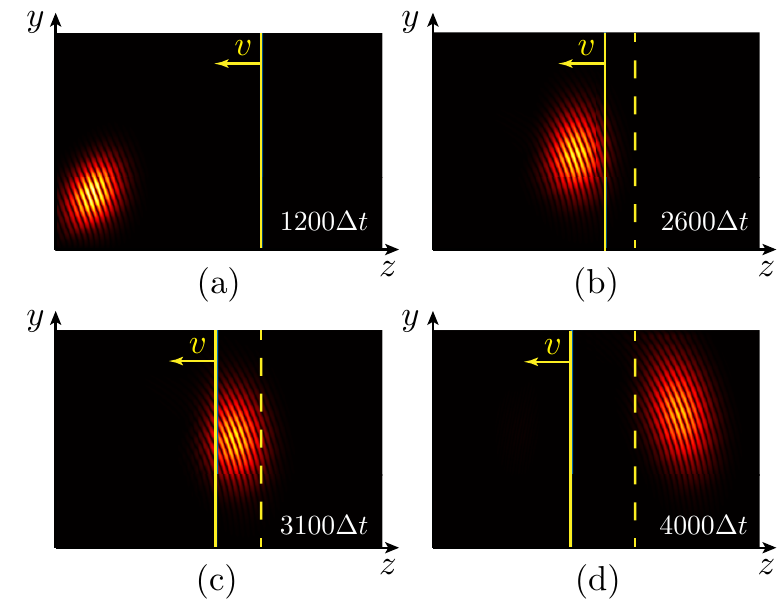}
    \caption{Verification and illustration of the inherent impedance- and velocity-matching of the generalized Yee cell with the conventional Yee cell for a moving interface ($v=-0.3c$) between two identical media ($\eps=1.5$), showing temporal snapshots where the wave is (a)~in the incident conventional Yee cell region, (b)~across the input edge of the conventional to generalized Yee cell regions [left edge of the staircase band in Fig.~\ref{fig:Gen_FDTD_approach}(b)], (c)~across the output edge of that region [right edge of the staircase band in Fig.~\ref{fig:Gen_FDTD_approach}(b)], and (d)~in the output conventional Yee cell region.}\label{fig:App_Matching}
\end{figure}

\pati{Matching Test}
Figure~\ref{fig:App_Matching} provides a numerical verification and illustration of the double impedance- and velocity-matching of the generalized Yee cell with its conventional counterpart by simulating an imaginary interface, with immaterial, testing Yee cell transition, between two identical media. As expected, neither scattering nor frequency transformation occur in that simulation.

\section{Scattering from an Accelerated \\ Perturbation Interface}\label{sec:App_ASTEM}
\pati{Introduction}
We derive here expressions for the fields scattered by an accelerated perturbation interface for bench-marking the numerical results in Sec.~\ref{sec:acc_interf}. The interface is assumed to have constant proper (or comoving-frame) acceleration, $a'$, which corresponds to the Rindler space-time metric~\cite{rindler1960hyperbolic} via Einstein's gravity-acceleration equivalence principle~\cite{einstein1907equivalence}.

\pati{Forms of the Fields}
The incident (i), reflected (r) and transmitted (t) electric and magnetic fields have the general forms
\begin{subequations}\label{eq:Acc_ass_E}
\begin{equation}\label{eq:Acc_ass_Ei}
    E_{\mt{i},x}=A_\mt{i}f(k_\mt{i}z-\omega_\mt{i}t),
\end{equation}
\begin{equation}\label{eq:Acc_ass_Er}
    E_{\mt{r},x}=A_\mt{r}f(k_\mt{r}z+\omega_\mt{r}t),
\end{equation}
\begin{equation}\label{eq:Acc_ass_Et}
    E_{\mt{t},x}=A_\mt{t}f(k_\mt{t}z-\omega_\mt{t}t),
\end{equation}
\end{subequations}
and
\begin{subequations}\label{eq:Acc_ass_H}
\begin{equation}\label{eq:Acc_ass_Hi}
    H_{\mt{i},y}=\frac{A_\mt{i}}{\eta_1}f(k_\mt{i}z-\omega_\mt{i}t),
\end{equation}
\begin{equation}\label{eq:Acc_ass_Hr}
    H_{\mt{r},y}=-\frac{A_\mt{r}}{\eta_1}f(k_\mt{r}z+\omega_\mt{r}t),
\end{equation}
\begin{equation}\label{eq:Acc_ass_Ht}
    H_{\mt{t},y}=\frac{A_\mt{t}}{\eta_2}f(k_\mt{t}z-\omega_\mt{t}t),
\end{equation}
\end{subequations}
where $f(\cdot)$ represents an arbitrary waveform profile.

\pati{Resolution Strategy and General Transformation Formulas}
We shall use again the frame-hopping strategy~\cite{van2012relativity}, i.e., transform the fields~\eqref{eq:Acc_ass_E} and~\eqref{eq:Acc_ass_H} into their comoving-frame counterparts, apply stationary boundary conditions in the moving frame, and transform the resulting complete fields back to the laboratory frame. The corresponding frame transformations are generally given by the tensorial coordinate transformations formulas~\cite{tu2017differential}
\begin{subequations}\label{eq:Covariant_Transformations}
\begin{equation}\label{eq:CT_a}
	F_{\mu' \nu'}=\frac{\partial x^\rho}{\partial x^{\mu'}}\frac{\partial x^\sigma}{\partial x^{\nu'}}F_{\rho \sigma}
\end{equation}
and
\begin{equation}
	\left|\text{det}\left(\frac{\partial x^{\mu'}}{\partial x^{\nu}} \right) \right|G^{\mu' \nu'}=\frac{\partial x^{\mu'}}{\partial x^\rho}\frac{\partial x^{\nu'}}{\partial x^\sigma}G^{\rho \sigma},
\end{equation}
\end{subequations}
where, in the case of electromagnetics, $F_{\mu \nu}$ and $G^{\mu \nu}$ are the Faraday tensor and its dual~\cite{van2012relativity}
\begin{subequations}\label{eq:Def_F_G}
\begin{equation}\label{eq:Def_F}
	F_{\mu \nu}=
	\begin{bmatrix}
		0 & E_x & E_y & E_z \\
		-E_x & 0 & -cB_z & cB_y\\
		-E_y & cB_z & 0 & -cB_x\\
		-E_z & -cB_y & cB_x & 0
	\end{bmatrix}
\end{equation}
and
\begin{equation}\label{eq:Def_G}
	G^{\mu \nu}=
	\begin{bmatrix}
		0 & -cD_x & -cD_y & -cD_z\\
		cD_x & 0 & H_z & -H_y\\
		cD_y & -H_z & 0 & H_x\\
		cD_z & H_y & -H_x & 0
	\end{bmatrix}.
\end{equation}
\end{subequations}

\pati{Rindler Transformations}
In addition to the electromagnetic fields in Eqs.~\eqref{eq:Def_F_G}, the tensorial coordinate transformations in Eqs.~\eqref{eq:Covariant_Transformations} also require the space-time variable Rindler transformations~\cite{rindler1960hyperbolic}, or their more general Kottler-M{\o}ller version~\cite{moller1972theory} for accommodating nonzero initial velocities, viz.,
\begin{subequations}\label{eq:Rinder_Transformations}
\begin{equation}\label{eq:Rindler_T}
	ct=\frac{c^2}{a'}\sqrt{g_{00}}\sinh(\xi+\xi_0)-\frac{c^2}{a'}\sinh(\xi_0),
\end{equation}
\begin{equation}
	x=x',
\end{equation}
\begin{equation}
	~y=y',
\end{equation}
\begin{equation}\label{eq:Rindler_Z}
	z=\frac{c^2}{a'}\sqrt{g_{00}}\cosh(\xi+\xi_0)-\frac{c^2}{a'}\cosh(\xi_0),
\end{equation}
where
\begin{equation}\label{eq:xit}
	\xi=a't'/c
\end{equation}
and
\begin{equation}\label{eq:xi0}
	\xi_0=\sinh^{-1}(\beta_0\gamma_0),
\end{equation}
with $\beta_0$ and $\gamma_0$ being the initial relative velocity and Lorentz factor, respectively, and
\begin{equation}\label{eq:Rinder_Metric_g}
	g_{00}=\left[1+(a'z'/c^2)\right]^2,
\end{equation}
\end{subequations}
and $g_{00}$ being the $00$-term of the metric $g_{\mu'\nu'}=\mathrm{diag}(g_{00},1,1,1)$  

\pati{Transformed Fields}
Substituting Eqs.~\eqref{eq:Acc_ass_E} and~\eqref{eq:Acc_ass_H} into Eqs.~\eqref{eq:Def_F_G} as well as Eqs.~\eqref{eq:Rinder_Transformations} into Eqs.~\eqref{eq:Covariant_Transformations} yields the transformed-field expressions
\begin{subequations}\label{eq:Acc_K'}
\begin{equation}\label{eq:Acc_K'_Ei}
    E'_{\mt{i},x}=A_\mt{i}\sqrt{g_{00}}\left[\cosh(\xi+\xi_0)-n_1\sinh(\xi+\xi_0)\right]f(\phi_\mt{i}'),
\end{equation}
\begin{equation}\label{eq:Acc_K'_Er}
    E'_{\mt{r},x}=A_\mt{r}\sqrt{g_{00}}\left[\cosh(\xi+\xi_0)+n_1\sinh(\xi+\xi_0)\right] f(\phi_\mt{r}'),
\end{equation}
\begin{equation}\label{eq:Acc_K'_Et}
    E'_{\mt{t},x}=A_\mt{t}\sqrt{g_{00}}\left[\cosh(\xi+\xi_0)-n_2\sinh(\xi+\xi_0)\right]f(\phi_\mt{t}')
\end{equation}
and
\begin{equation}\label{eq:Acc_K'_Hi}
    H'_{\mt{i},y}=\frac{A_\mt{i}}{\eta_1}\sqrt{g_{00}}\left[\cosh(\xi+\xi_0)-n_1\sinh(\xi+\xi_0)\right]f(\phi_\mt{i}'),
\end{equation}
\begin{equation}\label{eq:Acc_K'_Hr}
    H'_{\mt{r},y}=\frac{A_\mt{r}}{\eta_1}\sqrt{g_{00}}\left[\cosh(\xi+\xi_0)+n_1\sinh(\xi+\xi_0)\right]f(\phi_\mt{r}'),
\end{equation}
\begin{equation}\label{eq:Acc_K'_Ht}
    H'_{\mt{t},y}=\frac{A_\mt{t}}{\eta_2}\sqrt{g_{00}}\left[\cosh(\xi+\xi_0)-n_2\sinh(\xi+\xi_0)\right]f(\phi_\mt{t}'),
\end{equation}
\end{subequations}
where
\begin{subequations}
\begin{equation}
    \begin{split}
    \phi_\mt{i}'=&k_\mt{i}\left[\frac{c^2}{a'}\sqrt{g_{00}}\cosh(\xi+\xi_0)-\frac{c^2}{a'}\cosh(\xi_0)\right]
    \\&-\omega_\mt{i}\left[\frac{c}{a'}\sqrt{g_{00}}\sinh(\xi+\xi_0)-\frac{c}{a'}\sinh(\xi_0)\right],
    \end{split}
\end{equation}
\begin{equation}
    \begin{split}
    \phi_\mt{r}'=&k_\mt{r}\left[\frac{c^2}{a'}\sqrt{g_{00}}\cosh(\xi+\xi_0)-\frac{c^2}{a'}\cosh(\xi_0)\right]
    \\&+\omega_\mt{r}\left[\frac{c}{a'}\sqrt{g_{00}}\sinh(\xi+\xi_0)-\frac{c}{a'}\sinh(\xi_0)\right],
    \end{split}
\end{equation} 
\begin{equation}
    \begin{split}
    \phi_\mt{t}'=&k_\mt{t}\left[\frac{c^2}{a'}\sqrt{g_{00}}\cosh(\xi+\xi_0)-\frac{c^2}{a'}\cosh(\xi_0)\right]
    \\&-\omega_\mt{t}\left[\frac{c}{a'}\sqrt{g_{00}}\sinh(\xi+\xi_0)-\frac{c}{a'}\sinh(\xi_0)\right].
    \end{split}
\end{equation}
\end{subequations}
\pati{Scattered Electric-Field Formulas}
Enforcing the continuity of the tangential electric and magnetic fields across the comoving stationary interface and inverse-Rindler transforming the resulting expressions using~\eqref{eq:Covariant_Transformations} leads finally to the electric scattered fields
\begin{subequations}\label{eq:ASTEM_Scattered}
    \begin{equation}\label{eq:ASTEM_Scattered_ref}
    \begin{split}
        E_\mt{r}=&\frac{\eta_2-\eta_1}{\eta_2+\eta_1}\frac{1-n_1\tanh(\xi+\xi_0)}{1+n_1\tanh(\xi+\xi_0)}A_\mt{i}\\&\cdot f\left[\frac{1-n_1\tanh(\xi+\xi_0)}{1+n_1\tanh(\xi+\xi_0)}(k_\mt{i}z+\omega_\mt{i}t)\right] 
    \end{split}
    \end{equation}
    and
    \begin{equation}\label{eq:ASTEM_Scattered_trans}
    \begin{split}
        E_\mt{t}=&\frac{2\eta_2}{\eta_2+\eta_1}\frac{1-n_1\tanh(\xi+\xi_0)}{1-n_2\tanh(\xi+\xi_0)}A_\mt{i}\\&\cdot f\left[\frac{1-n_1\tanh(\xi+\xi_0)}{1-n_2\tanh(\xi+\xi_0)}(k_\mt{i}z+\omega_\mt{i}t)\right],
    \end{split}
    \end{equation}
\end{subequations}
where $\xi=\xi(t')$ and $\xi_0$ were given by Eqs.~\eqref{eq:xit} and~\eqref{eq:xi0}. Note that these relations are expressed here in terms of a combination of comoving- and laboratory-frame quantities for the sake of compactness; the complete laboratory-frame quantities are obtained by substitution of the inverse relations of Eqs.~\eqref{eq:Rindler_T} and~\eqref{eq:Rindler_Z}.

\ifCLASSOPTIONcaptionsoff
  \newpage
\fi


{\small
\bibliographystyle{IEEEtran}
\bibliography{FDTD_TAP_REF.bib}

\providecommand{\noopsort}[1]{}\providecommand{\singleletter}[1]{#1}%
\begin{thebibliography}{10}
\providecommand{\url}[1]{#1}
\csname url@samestyle\endcsname
\providecommand{\newblock}{\relax}
\providecommand{\bibinfo}[2]{#2}
\providecommand{\BIBentrySTDinterwordspacing}{\spaceskip=0pt\relax}
\providecommand{\BIBentryALTinterwordstretchfactor}{4}
\providecommand{\BIBentryALTinterwordspacing}{\spaceskip=\fontdimen2\font plus
\BIBentryALTinterwordstretchfactor\fontdimen3\font minus \fontdimen4\font\relax}
\providecommand{\BIBforeignlanguage}[2]{{%
\expandafter\ifx\csname l@#1\endcsname\relax
\typeout{** WARNING: IEEEtran.bst: No hyphenation pattern has been}%
\typeout{** loaded for the language `#1'. Using the pattern for}%
\typeout{** the default language instead.}%
\else
\language=\csname l@#1\endcsname
\fi
#2}}
\providecommand{\BIBdecl}{\relax}
\BIBdecl

\bibitem{deck2022yeecell}
Z.-L. Deck-L{\'e}ger, A.~Bahrami, Z.~Li, and C.~Caloz, ``Generalized {FDTD} scheme for the simulation of electromagnetic scattering in moving structures,'' \emph{Opt. Express}, vol.~31, no.~14, pp. 23\,214--23\,228, 2023.

\bibitem{caloz2022gstem}
C.~Caloz, Z.-L. Deck-L{\'e}ger, A.~Bahrami, O.~C. Vicente, and Z.~Li, ``Generalized space-time engineered modulation ({GSTEM}) metamaterials: {A} global and extended perspective.'' \emph{IEEE Antennas Propag. Mag.}, pp. 2--12, 2022.

\bibitem{bradley1729iv}
J.~Bradley, ``I{V}. {A} letter from the {R}everend {M}r. {J}ames {B}radley {S}avilian {P}rofessor of {A}stronomy at {O}xford, and {F}{R}{S} to {D}r. {E}dmond {H}alley {A}stronom. {R}eg. \&c. giving an account of a new discovered motion of the fixed stars,'' \emph{Philos. Trans. R. Soc.}, vol.~35, no. 406, pp. 637--661, 1729.

\bibitem{Fresnel1818}
A.~Fresnel, ``Lettre d'{A}ugustin {F}resnel \`{a} {F}ran\c{c}ois {A}rago sur l’influence du mouvement terrestre dans quelques ph\'{e}nom\`{e}nes d’optiques,'' \emph{Ann. Chim. Phys}, vol.~9, pp. 57--66, 1818.

\bibitem{doppler1903ueber}
C.~Doppler, ``\"{U}ber das farbige {L}icht der {D}oppelsterne und einiger anderer {G}estirne des {H}immels,'' \emph{K{\"o}nigl. B{\"o}hm Gedsellsch. d. {W}is.}, vol.~2, pp. 465--482, 1842.

\bibitem{Fizeau1851}
H.~Fizeau, ``Sur les hypothèses relatives à l’éther lumineux, et sur une expérience qui parait démontrer que le mouvement des corps change la vitesse avec laquelle la lumière se propage dans leur intérieur,'' \emph{CR Acad. Sci.}, vol.~33, pp. 349--355, 1851.

\bibitem{Rontgen_1888}
W.~C. R{\"o}ntgen, ``Ueber die durch {B}ewegung eines im homogenen electrischen {F}elde befindlichen {D}ielectricums hervorgerufene electrodynamische {K}raft,'' \emph{Ann. Phys.}, vol. 271, no.~10, pp. 264--270, 1888.

\bibitem{Minkowski_1908}
H.~Minkowski, ``Die {G}rundgleichungen f\"{u}r die elektromagnetischen {V}org\"{a}nge in bewegten k\"{o}rpern,'' \emph{Nachr. Ges. Wiss. Goettingen, Math.-Phys. Kl}, vol.~10, pp. 53--111, 1908.

\bibitem{tien1958parametric}
P.~K. Tien, ``Parametric amplification and frequency mixing in propagating circuits,'' \emph{J. Appl. Phys.}, vol.~29, no.~9, pp. 1347--1357, 1958.

\bibitem{cassedy1963dispersion}
E.~S. Cassedy and A.~A. Oliner, ``Dispersion relations in time-space periodic media: Part {I}—stable interactions,'' \emph{Proc. IEEE}, vol.~51, no.~10, pp. 1342--1359, 1963.

\bibitem{yeh1965}
C.~Yeh, ``Reflection and transmission of electrodynamic waves by a moving dielectric medium,'' \emph{J. Appl. Phys.}, vol.~36, no.~11, pp. 3513--3516, 1965.

\bibitem{cassedy1967dispersion}
E.~S. Cassedy, ``Dispersion relations in time-space periodic media part {II}—unstable interactions,'' \emph{Proc. IEEE}, vol.~55, no.~7, pp. 1154--1168, 1967.

\bibitem{kong_1968_wave}
J.-A. Kong and D.~K. Cheng, ``Wave behavior at an interface of a semi-infinite moving anisotropic medium,'' \emph{J. Appl. Phys.}, vol.~39, no.~5, pp. 2282--2286, 1968.

\bibitem{tanaka1978relativistic}
K.~Tanaka, ``Relativistic study of electromagnetic waves in the accelerated dielectric medium,'' \emph{J. Appl. Phys.}, vol.~49, no.~8, pp. 4311--4319, 1978.

\bibitem{leonhardt1999optics}
U.~Leonhardt and P.~Piwnicki, ``Optics of nonuniformly moving media,'' \emph{Phys. Rev. A}, vol.~60, no.~6, p. 4301, 1999.

\bibitem{Reed_2003_color}
E.~J. Reed, M.~Solja\ifmmode \check{c}\else \v{c}\fi{}i\ifmmode~\acute{c}\else \'{c}\fi{}, and J.~D. Joannopoulos, ``Color of shock waves in photonic crystals,'' \emph{Phys. Rev. Lett.}, vol.~90, p. 203904, May 2003.

\bibitem{Lurie_Springer_2007}
K.~A. Lurie, \emph{An Introduction to the Mathematical Theory of Dynamic Materials}.\hskip 1em plus 0.5em minus 0.4em\relax Springer, 2007.

\bibitem{Biancalana_2007_dynamics}
F.~Biancalana, A.~Amann, A.~V. Uskov, and E.~P. O'Reilly, ``Dynamics of light propagation in spatiotemporal dielectric structures,'' \emph{Phys. Rev. E}, vol.~75, p. 046607, Apr 2007.

\bibitem{philbin_2008_fiber}
T.~G. Philbin, C.~Kuklewicz, S.~Robertson, S.~Hill, F.~Konig, and U.~Leonhardt, ``Fiber-optical analog of the event horizon,'' \emph{Science}, vol. 319, no. 5868, pp. 1367--1370, 2008.

\bibitem{Yu_2009_opticalisolation}
Z.~Yu and S.~Fan, ``Complete optical isolation created by indirect interband photonic transitions,'' \emph{Nat. Photonics}, vol.~3, no.~2, pp. 91--94, 2009.

\bibitem{belgiorno_2010_hawking}
F.~Belgiorno, S.~L. Cacciatori, M.~Clerici, V.~Gorini, G.~Ortenzi, L.~Rizzi, E.~Rubino, V.~G. Sala, and D.~Faccio, ``Hawking radiation from ultrashort laser pulse filaments,'' \emph{Phys. Rev. Lett.}, vol. 105, no.~20, p. 203901, 2010.

\bibitem{faccio2012optical}
D.~Faccio, T.~Arane, M.~Lamperti, and U.~Leonhardt, ``Optical black hole lasers,'' \emph{Class. Quantum Grav.}, vol.~29, no.~22, p. 224009, 2012.

\bibitem{faccio2013analogue}
D.~Faccio, F.~Belgiorno, S.~Cacciatori, V.~Gorini, S.~Liberati, and U.~Moschella, \emph{Analogue {G}ravity {P}henomenology: {A}nalogue {S}pacetimes and {H}orizons, from {T}heory to {E}xperiment}.\hskip 1em plus 0.5em minus 0.4em\relax Springer, 2013, vol. 870.

\bibitem{Deck_2018_wave}
Z.-L. Deck-L{\'e}ger, A.~Akbarzadeh, and C.~Caloz, ``Wave deflection and shifted refocusing in a medium modulated by a superluminal rectangular pulse,'' \emph{Phys. Rev. B}, vol.~97, no.~10, p. 104305, 2018.

\bibitem{Shaltout_Science_2019}
A.~M. Shaltout, V.~M. Shalaev, and M.~L. Brongersma, ``Spatiotemporal light control with active metasurfaces,'' \emph{Science}, vol. 364, no. 6441, pp. 1--10, May 2019.

\bibitem{Deck_APH_10_2019}
Z.-L. Deck-L\'{e}ger, N.~Chamanara, M.~Skorobogatiy, M.~G. Silveirinha, and C.~Caloz, ``Uniform-velocity spacetime crystals,'' \emph{Adv. Photon.}, vol.~1, no.~5, pp. 056\,002:1--26, Oct. 2019, invited.

\bibitem{caloz2019spacetime1}
C.~Caloz and Z.-L. Deck-L{\'e}ger, ``Spacetime metamaterials—{P}art {I}: {G}eneral concepts,'' \emph{IEEE Trans. Antennas Propag.}, vol.~68, no.~3, pp. 1569--1582, 2019.

\bibitem{caloz2019spacetime2}
C.~{}Caloz and Z.-L. Deck-L{\'e}ger, ``Spacetime metamaterials—{P}art {II}: {T}heory and applications,'' \emph{IEEE Trans. Antennas Propag.}, vol.~68, no.~3, pp. 1583--1598, 2019.

\bibitem{huidobro2019fresnel}
P.~A. Huidobro, E.~Galiffi, S.~Guenneau, R.~V. Craster, and J.~Pendry, ``Fresnel drag in space--time-modulated metamaterials,'' \emph{Proc. Natl. Acad. Sci.}, vol. 116, no.~50, pp. 24\,943--24\,948, 2019.

\bibitem{Gaafar_2019_front}
M.~A. Gaafar, T.~Baba, M.~Eich, and A.~Y. Petrov, ``Front-induced transitions,'' \emph{Nat. Photonics}, vol.~13, no.~11, pp. 737--748, 2019.

\bibitem{huidobro2021homogenization}
P.~A. Huidobro, M.~G. Silveirinha, E.~Galiffi, and J.~Pendry, ``Homogenization theory of space-time metamaterials,'' \emph{Phys. Rev. Appl.}, vol.~16, no.~1, p. 014044, 2021.

\bibitem{Li_PRB_03_2023}
Z.~Li, X.~Ma, A.~Bahrami, Z.-L. Deck-L\'{e}ger, and C.~Caloz, ``Generalized total internal reflection at dynamic interfaces,'' \emph{Phys. Rev. B}, vol. 107, no.~11, pp. 115\,129:1--8, Mar. 2023.

\bibitem{bahrami2023astem}
A.~Bahrami, Z.-L. Deck-L\'eger, and C.~Caloz, ``Electrodynamics of accelerated-modulation space-time metamaterials,'' \emph{Phys. Rev. Appl.}, vol.~19, p. 054044, 2023.

\bibitem{yee1966numerical}
K.~Yee, ``Numerical solution of initial boundary value problems involving {M}axwell's equations in isotropic media,'' \emph{IEEE Trans. Antennas Propag.}, vol.~14, no.~3, pp. 302--307, 1966.

\bibitem{taflove2005computational}
A.~Taflove, S.~C. Hagness, and M.~Piket-May, \emph{Computational Electromagnetics: The Finite-Difference Time-Domain Method}.\hskip 1em plus 0.5em minus 0.4em\relax Elsevier Amsterdam, Netherlands, 2005, vol.~3.

\bibitem{Harfoush1990}
F.~Harfoush, A.~Taflove, and G.~A. Kriegsmann, ``Numerical implementation of relativistic electromagnetic boundary conditions in a laboratory-frame grid,'' \emph{J. Comput. Phys.}, vol.~89, no.~1, pp. 80--94, 1990.

\bibitem{Harfoush1989}
F.~Harfoush{}, A.~Taflove, and G.~A. Kriegsmann, ``A numerical technique for analyzing electromagnetic wave scattering from moving surfaces in one and two dimensions,'' \emph{IEEE Trans. Ant. Propag.}, vol.~37, no.~1, pp. 55--63, 1989.

\bibitem{iwamatsu2009}
H.~Iwamatsu and M.~Kuroda, ``Over set grid generation method coupled with {FDTD} method while considering the {D}oppler effect,'' \emph{IEEJ Trans. Fundam. Mater.}, vol. 129, no.~10, pp. 699--703, 2009.

\bibitem{zheng2016}
K.~Zheng, Z.~Mu, H.~Luo, and G.~Wei, ``Electromagnetic properties from moving dielectric in high speed with {L}orentz-{FDTD},'' \emph{IEEE Ant. Propag. Lett.}, vol.~15, pp. 934--937, 2016.

\bibitem{zhao_2018}
Y.~Zhao and S.~Chaimool, ``Relativistic {F}inite-{D}ifference {T}ime-{D}omain analysis of high-speed moving metamaterials,'' \emph{Sci. Rep.}, vol.~8, pp. 7686:1--12, 2018.

\bibitem{Bahrami_MTM_09_2022}
A.~Bahrami and C.~Caloz, ``{FDTD} scheme for interfaces formed by space-time modulations,'' in \emph{Metamaterials 2022}, Siena, Sep. 2022.

\bibitem{mur1981absorbing}
G.~Mur, ``Absorbing boundary conditions for the finite-difference approximation of the time-domain electromagnetic-field equations,'' \emph{IEEE Trans. Electromagn. Compat.}, no.~4, pp. 377--382, 1981.

\bibitem{berenger1993}
J.-P. Berenger, ``A perfectly matched layer for the absorption of electromagnetic waves,'' \emph{J. Comput. Phys.}, vol. 114, pp. 185--200, 1993.

\bibitem{Jackson_1998}
J.~D. Jackson, \emph{Classical Electrodynamics}, 3rd~ed.\hskip 1em plus 0.5em minus 0.4em\relax Wiley, 1998.

\bibitem{Pena_2020_tempcoating}
V.~Pacheco-Pe{\~n}a and N.~Engheta, ``Antireflection temporal coatings,'' \emph{Optica}, vol.~7, no.~4, pp. 323--331, Apr 2020.

\bibitem{Pacheco_2020_temporalaming}
V.~{}Pacheco-Pe{\~n}a and N.~Engheta, ``Temporal aiming,'' \emph{Light Sci. Appl.}, vol.~9, no.~1, p. 129, 2020.

\bibitem{pauli1981theory}
W.~Pauli, \emph{Theory of Relativity}.\hskip 1em plus 0.5em minus 0.4em\relax Courier Corporation, 1981.

\bibitem{bladel1984}
J.~V. Bladel, \emph{Relativity and {E}ngineering}.\hskip 1em plus 0.5em minus 0.4em\relax Springer Science, 1984.

\bibitem{kong2008theory}
J.~A. Kong, \emph{Electromagnetic Wave Theory}, 7th~ed.\hskip 1em plus 0.5em minus 0.4em\relax EMW Publication, 2008.

\bibitem{Deck_Photon_04_2021}
Z.-L. Deck-L\'{e}ger, X.~Zheng, and C.~Caloz, ``Electromagnetic wave scattering from a moving medium with stationary interface across the interluminal regime,'' \emph{Photonics}, vol.~8, no.~6, pp. 1--15, Jun. 2021.

\bibitem{balazs_1961_solution}
N.~L. Balazs, ``On the solution of the wave equation with moving boundaries,'' \emph{J. Math. Anal. Appl.}, vol.~3, no.~3, pp. 472--484, 1961.

\bibitem{Felsen_1970_wave}
L.~Felsen and G.~Whitman, ``Wave propagation in time-varying media,'' \emph{IEEE Trans. Antennas Propag.}, vol.~18, no.~2, pp. 242--253, 1970.

\bibitem{fante1971transmission}
R.~Fante, ``Transmission of electromagnetic waves into time-varying media,'' \emph{IEEE Trans. Antennas Propag.}, vol.~19, no.~3, pp. 417--424, 1971.

\bibitem{vonneumann1949}
J.~{von Neumann} and R.~D. Richtmyer, ``A method for the numerical calculation of hydrodynamic shocks,'' \emph{J. Appl. Phys}, vol.~21, no.~3, pp. 232--237, 1950.

\bibitem{Kunz_1980_plane}
K.~S. Kunz, ``Plane electromagnetic waves in moving media and reflections from moving interfaces,'' \emph{J. Appl. Phys.}, vol.~51, no.~2, pp. 873--884, 1980.

\bibitem{van2012relativity}
J.~Van~Bladel, \emph{Relativity and engineering}.\hskip 1em plus 0.5em minus 0.4em\relax Springer Science \& Business Media, 2012, vol.~15.

\bibitem{Morgenthaler_1958}
F.~R. Morgenthaler, ``Velocity modulation of electromagnetic waves,'' \emph{IRE Trans. Microw. Theory Tech.}, vol.~6, no.~2, pp. 167--172, Apr. 1958.

\bibitem{rindler1960hyperbolic}
W.~Rindler, ``Hyperbolic motion in curved space time,'' \emph{Phys. Rev.}, vol. 119, no.~6, p. 2082, 1960.

\bibitem{einstein1907equivalence}
A.~Einstein, ``\"{U}ber das {R}elativit\"{a}tsprinzip und die aus demselben gezogenen {F}olgerungen,'' \emph{Jahrbuch der Radioaktivit\"{a}t und Elektronik}, 1907.

\bibitem{tu2017differential}
L.~W. Tu, \emph{Differential {G}eometry: {C}onnections, {C}urvature, and {C}haracteristic {C}lasses}.\hskip 1em plus 0.5em minus 0.4em\relax Springer, 2017, vol. 275.

\bibitem{moller1972theory}
C.~M{\o}ller, \emph{The {T}heory of {R}elativity}.\hskip 1em plus 0.5em minus 0.4em\relax Clarendon Press, Oxford, 1972.

\end{thebibliography}
}

\end{document}